\documentclass[aps,pra,superscriptaddress,twocolumn,showpacs,floatfix,amsmath,amssymb]{revtex4-1}

\usepackage{dcolumn}              
\usepackage{bm}                   
\usepackage{graphicx}             
\usepackage{nicefrac}             
\usepackage[mathlines]{lineno}    
\makeatletter
\usepackage{color}

\usepackage{hyperref}
\hypersetup{
     colorlinks   = true,
     citecolor    = blue
}
\makeatother

\begin{document}

\title{Topological Magnon Insulator with a Kekul{\'e} Bond Modulation}

\author{Pierre A. Pantale{\'o}n}
\email{ppantaleon@uabc.edu.mx}
\affiliation{School of Physics and Astronomy, University of Manchester, Manchester M13 9PL, United Kingdom}

\author{ Ramon Carrillo-Bastos}
\affiliation{Facultad de Ciencias, Universidad Aut{\'o}noma de Baja California,
Apartado Postal 1880, 22800, Ensenada, Baja California, M{\'e}xico}

\author{Y. Xian}
\affiliation{School of Physics and Astronomy, University of Manchester, Manchester M13 9PL, United Kingdom}

\begin{abstract}
We examine the combined effects of a Kekul{\'e} coupling texture (KC) and a Dzyaloshinskii\textendash Moriya interaction (DMI) in a two\textendash dimensional ferromagnetic honeycomb lattice. By analyzing the gap closing conditions and the inversions of the bulk bands, we identify the parameter range in which the system behaves as a trivial or a nontrivial topological magnon insulator. We find four topological phases in terms of the KC parameter and the DMI strength. We present the bulk-edge correspondence for the magnons in a honeycomb lattice with an armchair or a zigzag boundary. Furthermore, we find Tamm-like edge states due to the intrinsic on-site interactions along the boundary sites. Our results may have significant implications to magnon transport properties in the 2D magnets at low temperatures.
\end{abstract}


\maketitle

\section{Introduction}

The investigation of topological insulators has revealed that the appearance of topologically protected edge states in a lattice with a boundary is a consequence of the nontrivial properties of the bulk energy bands. A well known example is the Kane-Mele model \cite{Kane2005}, where the spin-orbit coupling (SOC) in graphene causes a transition from a semi-metal to a quantum spin Hall insulator. Such transition is characterized by the formation of an insulating bulk gap and conducting gapless edge states which are robust against internal and external perturbations \cite{Delplace2011,Malki2017}. The robustness of the edge states is intimately related with the topological properties of the bulk bands. For example, in the quantum Hall and quantum spin Hall effects \cite{Laughlin1981,Halperin1982,Kane2005,Wu2006,Bernevig2006,Onoda2005}, the existence of gapless chiral edge states is found to be a consequence of bulk topological orders \cite{Thouless1982,Hatsugai1993,Hatsugai1993a,Qi2006}. 

Magnetic insulators, where the spin moments are carried by magnons, can also exhibit topological effects \cite{Onose2010,Madon2014,Mochizuki2014,Cao2015,Owerre2016c,Tanabe2016}. 
The magnon Hall effect has been observed in the ferromagnetic insulators $\mathrm{Lu}_{2}\mathrm{V}_{2}\mathrm{O}_{7}$, $\mathrm{Ho}_{2}\mathrm{V}_{2}\mathrm{O}_{7}$ and $\mathrm{In}_{2}\mathrm{Mn}_{2}\mathrm{O}_{7}$ \cite{Onose2010,Ideue2012}, in ferromagnetic crystals $\mathrm{Y}_{3}\mathrm{Fe}{}_{5}\mathrm{O}_{3}$~\cite{Madon2014,Tanabe2016}, in metal-organic kagom{\'e} magnets $\mathrm{Cu}$(1-3 bdc)~\cite{Hirschberger2015} and in the frustrated quantum magnet $\mathrm{Tb}_{2}\mathrm{Ti}_{2}\mathrm{O}_{7}$~\cite{Hirschberger2015a}; whereas a nontrivial topology of the bulk bands has been experimentally realized in both kagom{\'e}~\cite{Chisnell2015} and honeycomb~\cite{Chen2018} ferromagnetic lattices.  Theoretical studies have shown that by tuning the coupling parameters in a kagom{\'e} lattice \cite{Mook2014a,Mook2014, Seshadri2018}, a rich bulk topology emerges and, due to the bulk-edge correspondence, the thermal Hall effect and the number of edge states have been found to be dependent of the topological phases. Moreover, in the case of a honeycomb ferromagnetic lattice, it has been shown that the inclusion of a Dzyaloshinskii\textendash Moriya interaction (DMI) produces magnon edge states similar to those appearing in the Haldane model for spinless fermions \cite{Owerre2016d} and the Kane\textendash Mele model for electrons \cite{Kim2016a}. 

Most recently it has been revealed that bond modulations in two-dimensional fermionic systems may induce topological effects \cite{Amorim2016, Naumis2017, Gonzalez-Arraga2018}. In particular, it has been shown that different topological phases are obtained in graphene with a Kekul{\'e} bond modulation \cite{Kariyado2017,Liu2017} or the combined effect of  a Kekul{\'e} bond modulation with the SOC in the Kane-Mele model \cite{Grandi2015,Wu2016}. Because the topological physics is independent of the particle statistics, it will be interesting to investigate similar effects in magnetic lattices. Such is the case of ferromagnetic lattices where the strain induced modulations can be made and novel topological phases are thus obtained \cite{Ferreiros2018,Owerre2018}. 

In this paper, we report that different topological phases can be induced in a honeycomb ferromagnetic lattice by inclusion of bond modulations. More specifically, by exploring the combined effects of a Kekul{\'e} coupling modulation (KC) in the Heisenberg model and a Dzyaloshinskii\textendash Moriya interaction, we characterize four topological phases due to the band inversions and the Chern number of the bulk bands. Contrary to fermionic systems where the topological phases are characterized by the energy gap at Fermi level, we characterize the topological properties of 2D ferromagnetic systems by the  gaps at the low-lying energy bands. We identify the parameter range in which the system behaves as a trivial or a nontrivial topological magnon insulator. Furthermore, the bulk-edge correspondence for the edge magnons in a coupling modulated honeycomb ferromagnetic lattice is presented. In addition, we also find Tamm-like (trivial) edge states due to the intrinsic on-site interactions along the boundary sites \cite{Plotnik2014,Pantaleon2018a}. 

This paper is organized as follows: in Sec. \ref{sec: Bosonic Haldane model} we introduce the bosonic Haldane model with a KC texture and a DMI; in Sec. \ref{sec:Topological Phases} the gap closing conditions, band inversions and the topological phase diagram are obtained in terms of the coupling parameters.  In Sec. \ref{sec:Bulk Edge Correspondence}, by the bulk-edge correspondence, the edge states in an armchair and zigzag boundaries are characterized in each topological phase. Finally, the Sec. \ref{sec:Conclusions} is devoted to conclusions and discussions.

\section{The Bosonic Haldane model with a Kekul{\'e} Coupling Texture \label{sec: Bosonic Haldane model}}

\begin{figure}
\begin{centering}
\includegraphics[scale=0.70]{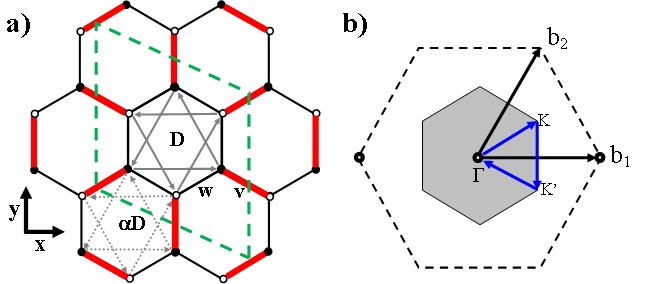}
\par\end{centering}
\caption{{\small{}(Color online) a) Lattice structure of a honeycomb ferromagnetic lattice with a KC texture and a DMI. The unit cell contains six sites with different values of second-nearest neighbor interactions as indicated. b) The Dirac points of the unperturbed Brillouin zone (dashed hexagon) are folded to the center of the new Brillouin zone (shaded region) due to the KC texture. The reciprocal lattice vectors are  $\boldsymbol b_{1}=\frac{2\pi}{3\sqrt{3}}\left(2,0\right)$ and $\boldsymbol b_{2}=\frac{2\pi}{3\sqrt{3}}\left(1,\sqrt{3}\right)$. \label{fig:Figure1}}}
\end{figure}

We consider the following Hamiltonian for a ferromagnetic honeycomb lattice: 
\begin{equation}
H=-\sum_{\left\langle i,j\right\rangle }J_{ij}\mathbf{S}_{i}\cdot\mathbf{S}_{j}+\sum_{\left\langle \left\langle i,j\right\rangle \right\rangle }\boldsymbol{D}_{ij}\cdot\left(\mathbf{S}_{i}\times\mathbf{S}_{j}\right),\label{eq: Spin Hamiltonian}
\end{equation}
 where the first summation runs over the nearest neighbors (NN) and the second runs over the next-nearest neighbors (NNN), $\mathbf{S}_{i}$ is the spin moment at site $i$ and, as shown in Fig. \ref{fig:Figure1}(a), in a KC texture the spins have a ferromagnetic (or exchange) coupling $J_{ij}=w$ if both $i,j$ belong to the same cell (intracell) and $J_{ij}=v$ otherwise (intercell). For a honeycomb lattice on the $xy$\textendash plane, the DMI vector $\boldsymbol{D}_{ij}=D_{ij}\varrho_{ij}\hat{z}$, where $\varrho_{ij}=\pm1$ is an orientation\textendash dependent coefficient in analogy with the Kane\textendash Mele model \cite{Kane2005}. Unlike benzene where the Kekul{\'e} ordering is due to the double carbon bond, the Kekul{\'e} distortion considered here accounts for bond modulations caused by local changes in the spin positions associated to local strain \cite{Ferreiros2018,Owerre2018}. In an isotropic lattice, the DMI strength is proportional to the exchange coupling parameter \cite{Moriya1960}. However, as shown in Fig.~\ref{fig:Figure1}(a), there are six sites in the unit cell and two values, $v$ and $w$, of ferromagnetic exchange parameters. Hence, as suggested in references \cite{Wu2012,Grandi2015}, we consider two values of the DMI strength: $D_{ij}=D$ and $D_{ij}=\alpha D$ with $\alpha=v/w$, for an intracell and intercell coupling, respectively. 
 
 By considering a ferromagnetic ground state and with the Holstein\textendash Primakoff transformations in the linear spin wave approximation \cite{Akhiezer1961}, the Hamiltonian in Eq. (\ref{eq: Spin Hamiltonian}) can be written as 
 
\begin{equation}
H=wS\sum_{\boldsymbol{k}}\Psi_{\boldsymbol{k}}^{\dagger}M_{\boldsymbol{k}}\Psi_{\boldsymbol{k}},\label{eq: Main Hamiltonian}
\end{equation}
where $\Psi_{k}$ is a $6$\textendash component vector, $S$ is the
spin quantum number and $M_{\boldsymbol{k}}$ a $6\times6$ matrix given by 
\begin{equation}
M_{\boldsymbol{k}}=M_{0}+M_{kek}+iD^{\prime}M_{D},\label{eq: Matrix M total Sum}
\end{equation}
with $M_{0}=(2+\alpha)I$ the on-site contribution, $I$ an identity matrix and $D^{\prime}=D/w$.  In the following, to simplify notation we omit the $\boldsymbol{k}$ dependence label. The additional terms of the above equation, 
\begin{align}
M_{kek} & =\left(\begin{array}{cc}
0 & -M_{1}\\
-M_{1}^{\dagger} & 0
\end{array}\right),\label{eq: Matrix M Kekule}\\
M_{D} & =\left(\begin{array}{cc}
M_{2} & 0\\
0 & M_{3}
\end{array}\right),\label{eq: Matrix M DMI}
\end{align}
are the matrices encoding the KC texture and the DMI, respectively, with the matrix elements given by, 

\begin{align}
M_{1} & =\left(\begin{array}{ccc}
\gamma_{1} & \alpha\gamma_{2} & \gamma_{3}\\
\gamma_{2} & \gamma_{3} & \alpha\gamma_{1}\\
\alpha\gamma_{3} & \gamma_{1} & \gamma_{2}
\end{array}\right),\nonumber \\
M_{2} & =\left(\begin{array}{ccc}
0 & -d_{1}^{\ast} & d_{2}\\
d_{1} & 0 & -d_{3}^{\ast}\\
-d_{2}^{\ast} & d_{3} & 0
\end{array}\right),\label{eq:Mij  Matrices}\\
M_{3} & =\left(\begin{array}{ccc}
0 & -d_{2} & d_{3}^{\ast}\\
d_{2}^{\ast} & 0 & -d_{1}\\
-d_{3} & d_{1}^{\ast} & 0
\end{array}\right),\nonumber 
\end{align}
where, 
\begin{align}
d_{1} & =\eta_{1}+\alpha\eta_{2}+\alpha\eta_{3},\nonumber \\
d_{2} & =\alpha\eta_{1}+\eta_{2}+\alpha\eta_{3},\label{eq: Mij matrix elements}\\
d_{3} & =\alpha\eta_{1}+\alpha\eta_{2}+\eta_{3}.\nonumber 
\end{align}
In the above equations, we have defined  that $\gamma_{i}~=~\exp\left(i \boldsymbol k\cdot \boldsymbol\sigma_{i}\right)$
and $\eta_{i}=\exp\left(i \boldsymbol k\cdot \boldsymbol \mu_{i}\right)$, with the sets given
by: $ \boldsymbol \sigma_{i}=\left\{ (\nicefrac{\sqrt{3}}{2},-\nicefrac{1}{2}),(0,1),(-\nicefrac{\sqrt{3}}{2},-\nicefrac{1}{2})\right\} $
and $\boldsymbol \mu_{i}=\left\{ (-\nicefrac{\sqrt{3}}{2},\nicefrac{3}{2}),(-\nicefrac{\sqrt{3}}{2},-\nicefrac{3}{2}),(\sqrt{3},0)\right\} $,
for the NN and NNN vectors respectively {[}See Fig. \ref{fig:Figure1}(a){]}. 
\begin{figure*}
\begin{centering}
\includegraphics[scale=0.55]{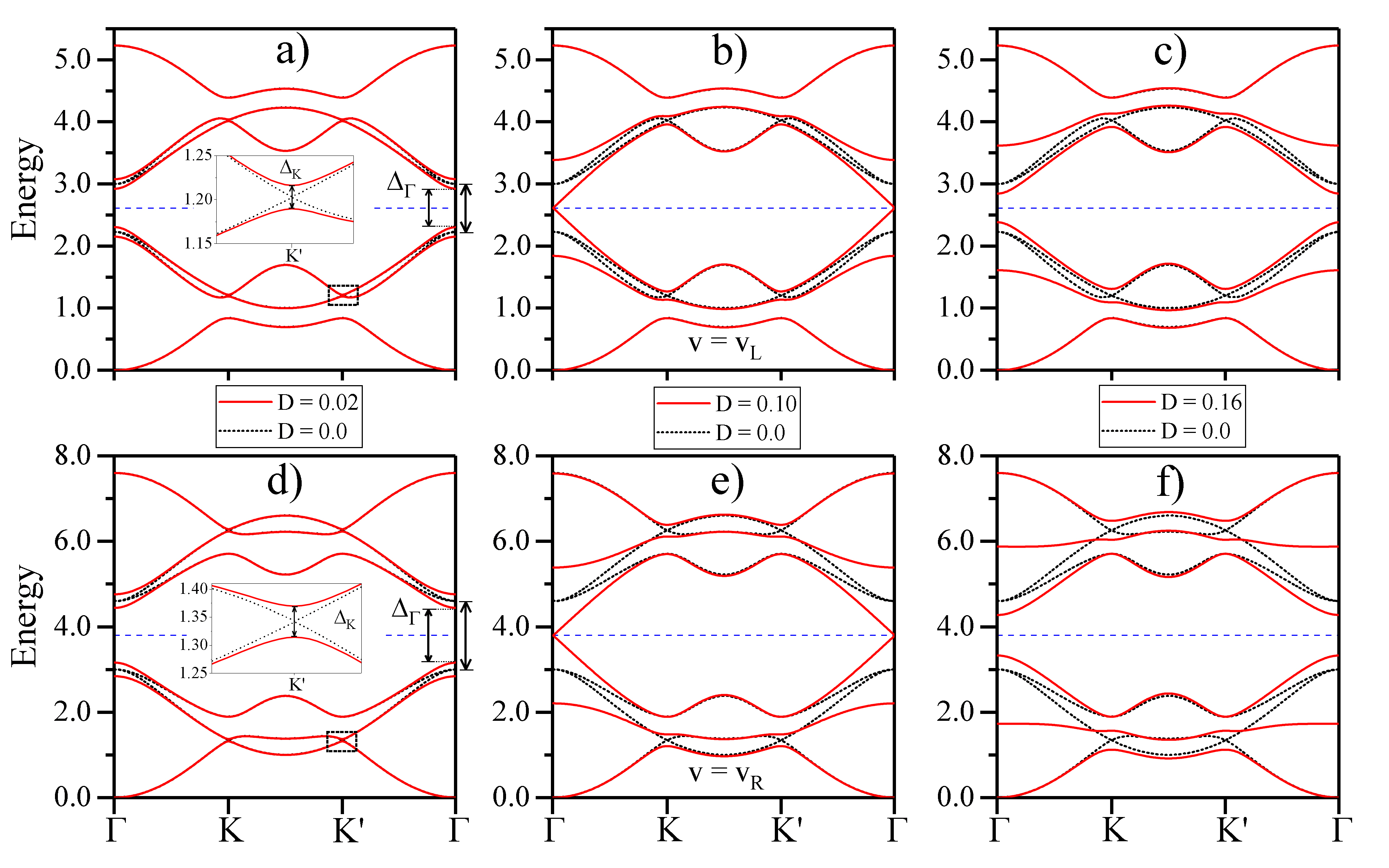}
\par\end{centering}
\caption{{\small{}(Color online) Energy band structure of a honeycomb ferromagnetic lattice with a KC texture and a DMI. In the set of figures, the KC parameters are: $v=0.614$ [(a)-(v)] and $v=1.8$ [(d)-(f)]. The (black) dotted lines are the band structure for $D=0$ while the (red) continuous lines are the band structure for different non-zero DMI values, namely: $D=0.02$  for (a) and (d), $D=0.1$ for (b) and (e) and $D=0.16$  for (c) and (f). The insets in (a) and (d) displays the enlarged region (dashed square) around the point $K^{\prime}$ in their respective panels. The dashed horizontal line in each panel is the corresponding bulk on-site energy $\varepsilon_{0}\left(=2w+v\right)$. \label{fig:Figure2}}}
\end{figure*}
If $v=w$, the Hamiltonian in Eq. (\ref{eq: Main Hamiltonian}) reduces to the bosonic Haldane model with a folded band structure \cite{Owerre2016d}. If $D=0$ and $v\neq w$, the system is similar to that of graphene with a bond-modulated honeycomb lattice, where two topological phases characterized by a change in the Zak phase have been identified \cite{Liu2017,Kariyado2017}. Although the Hamiltonian in Eq. (\ref{eq: Main Hamiltonian}) is similar to that of a fermionic lattice with SOC and a Kekul{\'e} bond modulation \cite{Grandi2015}, the bosonic nature of the magnons with the combined effect of the KC texture and the DMI, may provide us with additional topological phases and novel edge states. In the following, in all numerical calculations, the KC parameter $v$ and the DMI strength $D$ are given in the unit of $w(=1)$, the energy $\varepsilon$ is given in unit of $wS$. 

\section{Topological Phases \label{sec:Topological Phases}}

The presence of a Kekul{\'e} bond modulation between neighbouring sites increases the size of the original unit cell [dashed green line in Fig. \ref{fig:Figure1}(a)], hence, as shown in Fig. \ref{fig:Figure1}(b), the Dirac points from the original Brillouin zone are coupled and folded onto the center of the new reduced Brillouin zone \cite{Chamon2000,Hou2007,Gamayun2018}. For zero DMI, the Hamiltonian in Eq. (\ref{eq: Main Hamiltonian}) preserves both time reversal and sublattice symmetries. The characterization of their topological phases can be realized by the vectored Zak phase for the infinite system \cite{Liu2017} or by mirror winding numbers for the lattice with a boundary \cite{Kariyado2017}. A nonzero DMI comes from inversion symmetry breaking and results in the breaking of the effective time-reversal symmetry \cite{Owerre2016d, Mook2014,Pershoguba2017}. Hence, in terms of the KC parameter $v$ and the DMI strength $D$, novel topological phases are obtained. As shown in Figs. \ref{fig:Figure2}(a)\textendash (f), the eigenvalues of Eq. (\ref{eq: Main Hamiltonian}) are the six well-separated bulk bands, symmetrically located around the on-site energy $\varepsilon_{0}\left(=2w+v\right)$. Since a topological phase transition requires the energy gap closing down, we may then identify the gap closing  conditions in terms of the hopping parameters and the topological phases through the analysis of the band inversions \cite{Murakami2017}. We find that the analysis of the eigenvalues of Eq. (\ref{eq: Main Hamiltonian}) at the points $\Gamma\left(0,0\right)$ and $K\left(2\pi/3\sqrt{3},2\pi/9\right)$ are sufficient to determine the topological phase transitions in the system. 

\subsection{Gap closing conditions \label{sec:Gap Closing}}

We first consider the point $\Gamma$ where, from the six eigenvalues $\varepsilon_{1}=0$, $\varepsilon_{2}^{\pm}=\frac{1}{w}\left(2v+w\right)\left(\pm\sqrt{3}D+w\right)$,
$\varepsilon_{3}^{\pm}=\frac{1}{w}\left[3w^{2}\pm\left(2v+w\right)\sqrt{3}D\right]$
and $\varepsilon_{4}=2(v+2w)$, of Eq. (\ref{eq: Main Hamiltonian}), only $\varepsilon_{2}^{\pm}$ and $\varepsilon_{3}^{\pm}$ have their values around the on-site energy $\varepsilon_{0}$. Therefore, the energy gap at the point $\Gamma$ is given by

\begin{equation}
\Delta_{\varGamma}=\left|2\sqrt{3}D\left(1+2\frac{v}{w}\right)-2\left|w-v\right|\right|.\label{eq: Gap at Gamma}
\end{equation}
The energy gap closures occurs when $\Delta_{\varGamma}=0$, providing
the following critical values,
\begin{figure} \begin{centering} \includegraphics[scale=0.30]{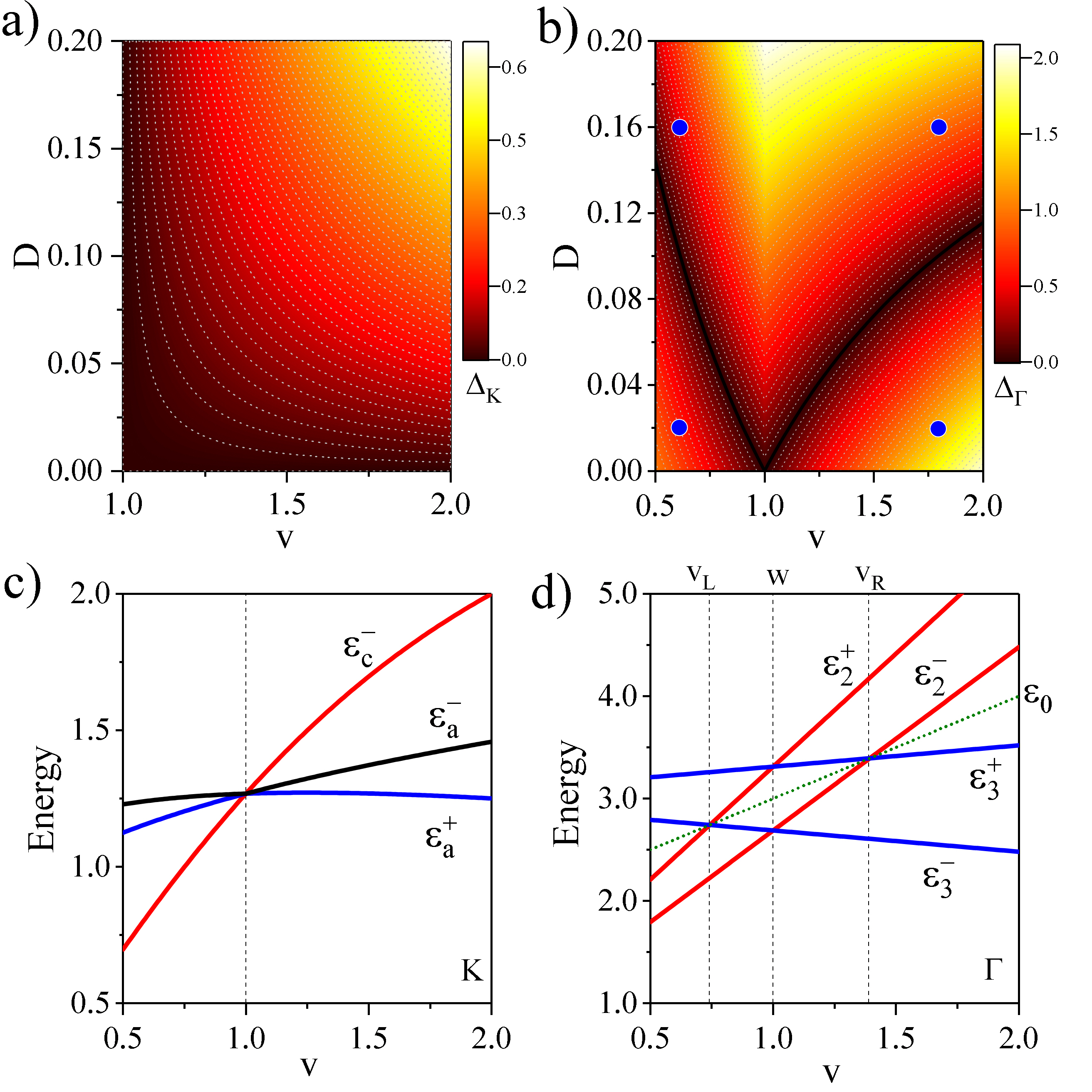} \par\end{centering}
\caption{{\small{}(Color online) Evolution of the energy gaps, a) $\Delta_{K}$ and b) $\Delta_{\varGamma}$, as a function of the KC parameter and the DMI strength. The (blue) circles mark those variable sets $\{v,\,D\}$ for which the band structure for the lattice with a boundary is calculated. c) Band evolution at the point $K$ for increasing $v$ and $D=0.06$ where a band inversion occurs at $v=w(=1)$. d) Band evolution at the point $\Gamma$ for increasing $v$ and a fixed $D=0.06$, where band inversions occur at the critical values: $v=v_{L}$, $v=w$ and $v=v_{R}$. The dotted (green) line is the bulk on-site energy $\varepsilon_{0}$. \label{fig:Figure3p1}}} \end{figure}
\begin{equation}
v_{L(R)}=\frac{w\left(6D^{2}\mp3\sqrt{3}Dw+w^{2}\right)}{w^{2}-12D^{2}},\label{eq: VL gap close down}
\end{equation}
for the KC parameter as a function of the DMI strength. In the above equation, the solutions $v_{L}(<w)$ and $v_{R}(>w)$ correspond to the negative and positive sign, respectively. In Figs. \ref{fig:Figure2}(a)\textendash (f) we display the energy bands along the path given by the arrows in Fig. \ref{fig:Figure1}(b), where for a given $v$ {[}$v=0.614$ (top row) and $v=1.80$ (bottom row){]} the DMI strength is varied. As mentioned before, for $D=0$ this system preserves both time reversal and sublattice symmetries. In such case, the energy gap, $\Delta_{\Gamma}$, in Eq. (\ref{eq: Gap at Gamma}) is proportional to the difference of the intracell ($w$) and intercell ($v$) coupling. The energy bands for $D=0$ are the (black) dotted lines in Figs. \ref{fig:Figure2}(a)\textendash (f), where $\Delta_{\varGamma}=0.77$ and $\Delta_{\varGamma}=1.6$, for the top and bottom rows, respectively. In addition, as shown in Figs. \ref{fig:Figure2}(b) and (e), for a given $D\left(=0.1\right)$, there are two critical values, $v_{L}(=0.614)$ and $v_{R}(=1.80)$, given by Eq. (\ref{eq: VL gap close down}), where $\Delta_{\varGamma}=0$. 

Now we consider the point $K(K^{\prime})$ in the reduced Brilloin zone, where as displayed in the insets of Figs. \ref{fig:Figure2}(a) and (d), the band structure is Dirac-like \cite{Fransson2016}. For a nonzero DMI, the magnons accumulate an additional phase upon propagation between NNN sites, therefore the degeneracy at the $K(K^{\prime})$ point is broken and an energy gap is thus induced. At the $K$ point, the  eigenvalues of Eq. (\ref{eq: Main Hamiltonian}), are given by $\varepsilon_{a}^{\pm}=\varepsilon_{0}-\frac{1}{w}\sqrt{3D^{2}v_{0}{}^{2}+w^{2}v_{1}^{2}\pm2\sqrt{3}Dw\left|v_{0}\right|v_{1}}$,
$\varepsilon_{b}^{\pm}=2\varepsilon_{0}-\varepsilon_{a}^{\pm}$ and $\varepsilon_{c}^{\pm}=\varepsilon_{0}\pm\sqrt{v^{2}-2vw+4w^{2}}$, with $v_{0}=v-w$ and $v_{1}^{2}=v^{2}+vw+w^{2}$. The energy gap, determined by the difference $\varepsilon_{a}^{-}-\varepsilon_{a}^{+}$, is trivial for $v<w$ and nontrivial for $v>w$, its explicit form is given by
\begin{equation} 
\Delta_{K}=2\sqrt{3}\left|\frac{v-w}{w}\right|D. \label{eq: Gap at K}
\end{equation}
The different panels in Fig. \ref{fig:Figure3p1} characterize the properties of the energy gaps, $\Delta_{K}$ and $\Delta_{\Gamma}$, as well as their dependence with the parameters $v$ and $D$. In Fig. \ref{fig:Figure3p1}(a), a plot of the energy gap $\Delta_{K}$ as a function of the parameters $v$ and $D$ is presented. We consider a range of parameters where $v\geq w$, such that [in agreement with Eq.(\ref{eq: Gap at K})] the value of the energy gap grows linearly with $D$. In Fig. \ref{fig:Figure3p1}(b), a black \textit{v\textendash shaped} line shows the critical values satisfying Eq. (\ref{eq: VL gap close down}), where $v_{L}(v_{R})$ is the line to the left(right) of the critical point $w=1$. Furthermore, for a given $D$, in the region $v<v_{L}$ or $v>v_{R}$ the energy gap, $\Delta_{\Gamma}$, increases as the $v$ value moves away from the critical values. In the complementary region ($v_{L}<v<v_{R}$), $\Delta_{\varGamma}$ approaches a maximum as $v\rightarrow w$, while the energy gap $\Delta_{K}\rightarrow0$. At this critical point, the system has an energy gap of $\Delta_{\varGamma}=6\sqrt{3}D$, as in the bosonic Haldane model \cite{Owerre2016d}. 

\begin{figure} \begin{centering} \includegraphics[scale=0.35]{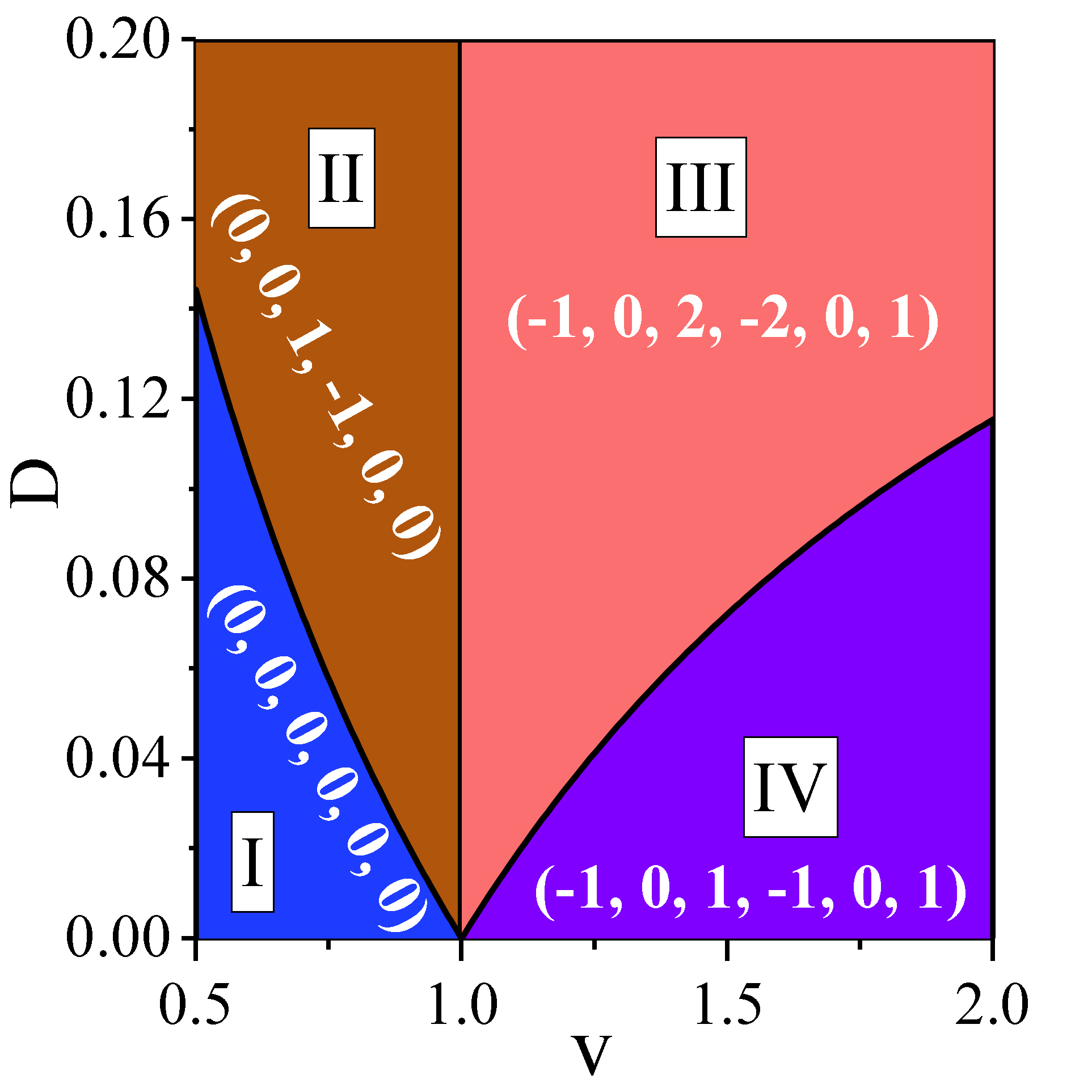} \par\end{centering}
\caption{{\small{}(Color online) Topological phase diagram of the bond-modulated honeycomb lattice with a Dzialoshinskii\textendash Moriya interaction, where each region is characterized by a set $\left(C_{1},C_{2}...,C_{6}\right)$ of Chern numbers.\label{fig:Figure3p2}}} \end{figure}

\begin{figure*}
\begin{centering}
\includegraphics[scale=0.4]{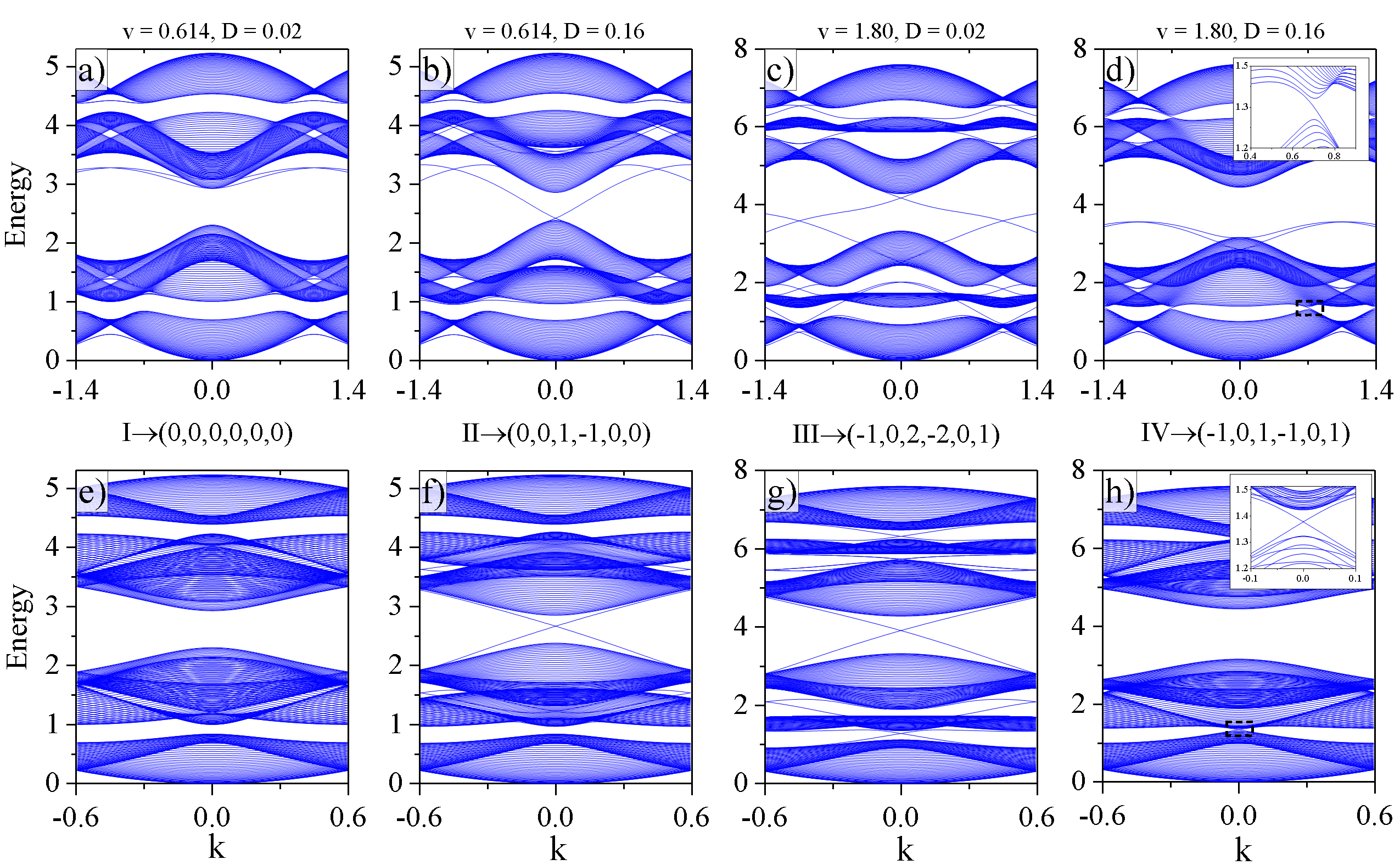}
\par\end{centering}
\caption{{\small{}(Color online) Energy band structure of a honeycomb lattice with armchair [(a)-(c)]) or zigzag [(d)-(f)] boundaries for different topological phases. The topological phases with their corresponding set of Chern numbers are indicated in each column [phase I (a) and (f), phase II (b) and (f), phase III ((v) and (g) and phase IV (d) and (h)]. The lines crossing the energy gaps and connecting adjacent bulk bands are the nontrivial edge modes, while the remaining are Tamm-like or trivial edge modes. The insets in d) and h) are the enlarged regions (dashed rectangle) with $\nu_{1}=-1$.  \label{fig:Figure4}}}
\end{figure*}

\subsection{Topological phase diagram \label{sec:Band inversions}}

Having identified the closing gap conditions in terms of the KC parameter and the DMI strength in the previous section, we proceed to identify the band inversions and the topological phases of the system. For a given $D$, and from Eqs. (\ref{eq: Gap at Gamma}) and (\ref{eq: Gap at K}), we have identified two critical points, $v_{L}$ and $v_{R}$ where $\Delta_{\Gamma}=0$, and a single point  $v=w$, where $\Delta_{K}=0$. Such critical points are associated with band inversions. At the point $K$, the band evolution for increasing $v$ and a given $D$ is shown in Fig. \ref{fig:Figure3p1}(c), where a band inversion  occurs at the critical value $v=w$. At the point $\Gamma$, Fig. \ref{fig:Figure3p1}(d), each energy band is inverted twice, however, by Eq. (\ref{eq: Gap at Gamma}), only the inversions at $v_{L}$ and $v_{R}$ are due to an energy gap closure. Thus, by the gap closing conditions and the band inversions, three phase boundaries are identify for this system. The resulting topological phase diagram is given in Fig. \ref{fig:Figure3p2}.  

On the other hand, as a result of the  DMI the magnons accumulate an additional phase upon propagation between NNN sites, giving rise to a six well defined bulk energy bands as shown in Fig. \ref{fig:Figure2}, with a non-vanishing Berry curvature. For a given values of the KC parameter and DMI strength, the Berry curvature of the $j$th band is given by~\cite{berry84},
\begin{equation}
\Omega_{j,\boldsymbol{k}}=i\sum_{j^{\prime}\neq j}\frac{\left\langle \psi_{j,\boldsymbol{k}}\left|\nabla_{\boldsymbol{k}}M_{\boldsymbol{k}}\right|\text{\ensuremath{\psi}}_{j^{\prime},\boldsymbol{k}}\right\rangle \times\left\langle \psi_{j^{\prime},\boldsymbol{k}}\left|\nabla_{\boldsymbol{k}}M_{\boldsymbol{k}}\right|\text{\ensuremath{\psi}}_{j,\boldsymbol{k}}\right\rangle }{\left(\varepsilon_{j,\boldsymbol{k}}-\varepsilon_{j^{\prime},\boldsymbol{k}}\right)^{2}},
\label{eq: Omega}
\end{equation}
where $\varepsilon_{j, \boldsymbol{k}}$ and $\psi_{j, \boldsymbol{k}}$ are the corresponding eigenvalues and eigenvectors of the Hamiltonian in Eq. (\ref{eq: Main Hamiltonian}). In analogy with a kagom{\'e} lattice \cite{Mook2014}, each phase in Fig.~\ref{fig:Figure3p2} is characterized by a set of Chern numbers $(C_{1},C_{2}\ldots,C_{6})$, where each $C_{j}$ is given by the integral of the Berry curvature over the Brillouin zone, that is
\begin{equation}
\label{eq:Chern}
C_j = \frac{1}{2 \pi} \int_\mathrm{BZ} d^2 k \, \Omega_{j,\boldsymbol{k}}.
\end{equation}
The Chern number in the above equation can be calculated by a direct numerical integration or with the plaquette method introduced in Ref. \cite{Fukui2005}. As shown in the topological phase diagram in Fig. \ref{fig:Figure3p2}, four phases in terms of the KC parameters and the DMI strength can be identified.  In contrast with the equivalent fermionic model \cite{Grandi2015}, the region $I$ in Fig. \ref{fig:Figure3p2}, with all zero Chern numbers, is the only trivial phase. The remaining phases, $II$, $III$ and $IV$, are nontrivial and they support topological edge states in a terminated lattice.
 
 \section{Bulk-Edge Correspondence \label{sec:Bulk Edge Correspondence}}
 
For the magnon excitations the number of topological edge states is related with the Chern number \cite{Hatsugai1993a,Hatsugai1993}. In analogy with a magnonic crystal \cite{Shindou2013} or a kagom{\'e} lattice \cite{Mook2014, Seshadri2018}, the winding number of the edge states in the band gap $i$, is given by 
\begin{equation}
\nu_{i}=\sum_{j\leq i}C_{j},\label{eq: Winding number}
\end{equation}
where $C_{j}$ is the Chern number of the band $j$. For the $i$th band gap, $\left|v_{i}\right|$ is the number of topological edge states and $sgn\left(v_{i}\right)$ their propagation direction. Since the emergence of edge states does not depend of the boundary type, we characterized the edge states in the different topological phases given in Fig. \ref{fig:Figure3p2}(e) for just zigzag and armchair boundaries. We investigate the edge states by considering boundary conditions following references \cite{Celic1996,You2008a,Sakaguchi2016}. In order to avoid finite size effects or energy gaps due to interference between edge states at opposite boundaries, a wide ribbon geometry is considered \cite{Ezawa2013}. The energy band structure is obtained for each topological phase with their corresponding set of parameters, \{$v,D$\}, indicated by (blue) circles in Fig. \ref{fig:Figure3p1}(b). In Fig. \ref{fig:Figure4}, the top and bottom panels display the energy band structure of a honeycomb lattice with armchair and zigzag boundaries, respectively. The lines crossing the energy gaps (connecting adjacent bulk bands) are topologically protected magnon edge states whereas the lines separated from the bulk bands (and not connecting adjacent bulk bands) are Tamm-like (or trivial) edge states \cite{Plotnik2014,Pantaleon2017a}. In contrast with the fermionic case, in a bosonic lattice the interaction terms along the outermost sites differ from the bulk values. Such a difference plays the role of an effective defect and gives rise to Tamm-like edge states. Lacking of topological protection, these trivial edge states are sensitives to external on-site potentials and they can be used to modify the dispersion relation of the nontrivial edge states \cite{Mook2014,Pantaleon2018a}. In the following we discuss the edge states in each topological phase separately. 

The phase $I$ in Fig. \ref{fig:Figure3p2}, with all zero Chern numbers, is the only trivial phase. As displayed in the band structure in Figs. \ref{fig:Figure4}(a) and (e), there are not lines crossing any of the energy gaps and no topological edge states are found. The lines separated from the bulk bands in Fig. \ref{fig:Figure4}(a), are the modes of Tamm-like edge states \cite{Tamm}. However, due to the Bose statistics, for a finite temperature all states contribute to transport and we expect a nonzero thermal magnon Hall conductivity in this phase. The details of the thermal magnon Hall conductivity for this system will be published elsewhere. In addition, we noticed that, in the zigzag boundary, Fig. \ref{fig:Figure4}(e), not flat bands are found, in agreement with previous results \cite{Pantaleon2018a}.  

In the phase $II$ and from Eq. (\ref{eq: Winding number}), the only nonzero winding number $\nu_{3}=1$ predicts a single edge state in the third energy gap. As displayed in the band structure in Figs. \ref{fig:Figure4}(b) and (f), the mode with positive slope crossing the third energy gap and connecting the bulk bands is the dispersion relation of the predicted nontrivial edge state at the upper boundary. The edge mode with negative slope is the energy spectrum of a nontrivial edge state at the opposite boundary. 

In the phase $III$ and as displayed in the band structure in Figs. \ref{fig:Figure4}(c) and (g), the nontrivial edge modes with $v_{i}=-1$, $(i=1,2,4,5)$, are the lines connecting the $(i+1)$th with the $i$th bulk bands. The nontrivial edge mode with winding number $v_{3}=1$, runs from the third to the fourth bulk band. For the opposite boundary, the nontrivial edge modes can be readily identified in the band structure due to the chirality of the magnon edge states \cite{Shindou2013}. 

 The phase $IV$ is shown in Figs. \ref{fig:Figure4}(d) and (h). In the equivalent fermionic model with zero energy at Fermi level \cite{Grandi2015}, phases $I$ and $IV$ are trivial. However, in a bosonic lattice at low temperatures, the edge modes in the first band gap are more populated than the edge modes with higher energy. Therefore, the phase $IV$ with winding numbers, $v_{1}=v_{2}=v_{4}=v_{5}=-~1$ and $v_{3}=0$, is a magnon nontrivial phase. We notice that the energy band structure has only three energy gaps and only two of them with nontrivial edge modes. This apparent inconsistency is due to an energy overlapping for the given values of parameters. The overlap can be removed to reveal the predicted edge states by modifying the KC parameter or DMI strength.

\section{Conclusions \label{sec:Conclusions}}

We have studied the topological phases and the emergence of edge states in a honeycomb ferromagnetic lattice with a Kekul{\'e} coupling texture and a Dzyaloshinskii\textendash Moriya interaction. By a bosonic tight binding model we have shown nontrivial topological phases in the 2D lattice system. The topological phases have been characterized in terms of the Kekul{\'e} coupling parameter and the Dzyaloshinskii\textendash Moriya interaction strength. In contrast to the fermionic case, we have found that the system has a single trivial and three nontrivial topological phases, characterized by the low-lying energy spectra and associated with a set of Chern numbers. These Chern numbers predict the number and propagation direction of the magnon edge states in a lattice with a boundary. We also find Tamm-like edge states due to the intrinsic on-site interactions along the boundary sites. We have presented the details of the energy spectra in the different topological phases, which are important for the investigation of the magnon transport in this system \cite{Mook2014,Mook2015}. 

Recently, the interesting nontrivial band structure of a ferromagnetic honeycomb lattice system chromium trihalides has been reported by Chen et al. \cite{Chen2018}, where the measured magnon spectrum is in agreement with the prediction of the Hamiltonian in Eq. (\ref{eq: Spin Hamiltonian}) without a KC modulation and for a nonzero DMI. While the model presented here does not correspond to the real system yet, it produces different topological phases in magnetic honeycomb lattices if the distance between magnetic moments is modulated by local strains \cite{Ferreiros2018,Owerre2018}. Similar to graphene where different Kekul{\'e} bond modulations can be induced by atoms adsorbed on its surface \cite{Cheianov2009a} or by a proximity with a substrate \cite{Gutierrez2016} it will be interesting to investigate the similar bond modulations in the magnetic lattices. Furthermore, the experimental discoveries of intrinsic 2D ferromagnetism in Van der Waals materials \cite{Gong2017,Huang2017,Miao2018} suggest that the edge magnon excitations may be realizable in a honeycomb ferromagnetic lattice \cite{Pershoguba2017}. Therefore, the characterization of the different topological phases and the edge states in 2D honeycomb ferromagnetic lattices presented in this paper may be useful for future experiments and magnonics applications \cite{Chumak2015,Chumak2017}. 

\section*{Acknowledgments}

We acknowledge useful discussions with Elias Andrade, Rory Brown and Christian Moulsdale. Pierre A. Pantale{\'o}n is sponsored by Mexico's National Council of Science and Technology (CONACYT) under scholarship No. 381939.

\noindent \bibliographystyle{apsrev4-1}
\bibliography{soc}

\begin{thebibliography}{65}%
\makeatletter
\providecommand \@ifxundefined [1]{%
 \@ifx{#1\undefined}
}%
\providecommand \@ifnum [1]{%
 \ifnum #1\expandafter \@firstoftwo
 \else \expandafter \@secondoftwo
 \fi
}%
\providecommand \@ifx [1]{%
 \ifx #1\expandafter \@firstoftwo
 \else \expandafter \@secondoftwo
 \fi
}%
\providecommand \natexlab [1]{#1}%
\providecommand \enquote  [1]{``#1''}%
\providecommand \bibnamefont  [1]{#1}%
\providecommand \bibfnamefont [1]{#1}%
\providecommand \citenamefont [1]{#1}%
\providecommand \href@noop [0]{\@secondoftwo}%
\providecommand \href [0]{\begingroup \@sanitize@url \@href}%
\providecommand \@href[1]{\@@startlink{#1}\@@href}%
\providecommand \@@href[1]{\endgroup#1\@@endlink}%
\providecommand \@sanitize@url [0]{\catcode `\\12\catcode `\$12\catcode
  `\&12\catcode `\#12\catcode `\^12\catcode `\_12\catcode `\%12\relax}%
\providecommand \@@startlink[1]{}%
\providecommand \@@endlink[0]{}%
\providecommand \url  [0]{\begingroup\@sanitize@url \@url }%
\providecommand \@url [1]{\endgroup\@href {#1}{\urlprefix }}%
\providecommand \urlprefix  [0]{URL }%
\providecommand \Eprint [0]{\href }%
\providecommand \doibase [0]{http://dx.doi.org/}%
\providecommand \selectlanguage [0]{\@gobble}%
\providecommand \bibinfo  [0]{\@secondoftwo}%
\providecommand \bibfield  [0]{\@secondoftwo}%
\providecommand \translation [1]{[#1]}%
\providecommand \BibitemOpen [0]{}%
\providecommand \bibitemStop [0]{}%
\providecommand \bibitemNoStop [0]{.\EOS\space}%
\providecommand \EOS [0]{\spacefactor3000\relax}%
\providecommand \BibitemShut  [1]{\csname bibitem#1\endcsname}%
\let\auto@bib@innerbib\@empty
\bibitem [{\citenamefont {Kane}\ and\ \citenamefont {Mele}(2005)}]{Kane2005}%
  \BibitemOpen
  \bibfield  {author} {\bibinfo {author} {\bibfnamefont {C.~L.}\ \bibnamefont
  {Kane}}\ and\ \bibinfo {author} {\bibfnamefont {E.~J.}\ \bibnamefont
  {Mele}},\ }\href {\doibase 10.1103/PhysRevLett.95.226801} {\bibfield
  {journal} {\bibinfo  {journal} {Phys. Rev. Lett.}\ }\textbf {\bibinfo
  {volume} {95}},\ \bibinfo {pages} {226801} (\bibinfo {year}
  {2005})}\BibitemShut {NoStop}%
\bibitem [{\citenamefont {Delplace}\ \emph {et~al.}(2011)\citenamefont
  {Delplace}, \citenamefont {Ullmo},\ and\ \citenamefont
  {Montambaux}}]{Delplace2011}%
  \BibitemOpen
  \bibfield  {author} {\bibinfo {author} {\bibfnamefont {P.}~\bibnamefont
  {Delplace}}, \bibinfo {author} {\bibfnamefont {D.}~\bibnamefont {Ullmo}}, \
  and\ \bibinfo {author} {\bibfnamefont {G.}~\bibnamefont {Montambaux}},\
  }\href {\doibase 10.1103/PhysRevB.84.195452} {\bibfield  {journal} {\bibinfo
  {journal} {Phys. Rev. B}\ }\textbf {\bibinfo {volume} {84}},\ \bibinfo
  {pages} {195452} (\bibinfo {year} {2011})}\BibitemShut {NoStop}%
\bibitem [{\citenamefont {Malki}\ and\ \citenamefont
  {Uhrig}(2017)}]{Malki2017}%
  \BibitemOpen
  \bibfield  {author} {\bibinfo {author} {\bibfnamefont {M.}~\bibnamefont
  {Malki}}\ and\ \bibinfo {author} {\bibfnamefont {G.~S.}\ \bibnamefont
  {Uhrig}},\ }\href {\doibase 10.1103/PhysRevB.95.235118} {\bibfield  {journal}
  {\bibinfo  {journal} {Phys. Rev. B}\ }\textbf {\bibinfo {volume} {95}},\
  \bibinfo {pages} {235118} (\bibinfo {year} {2017})}\BibitemShut {NoStop}%
\bibitem [{\citenamefont {Laughlin}(1981)}]{Laughlin1981}%
  \BibitemOpen
  \bibfield  {author} {\bibinfo {author} {\bibfnamefont {R.~B.}\ \bibnamefont
  {Laughlin}},\ }\href {\doibase 10.1103/PhysRevB.23.5632} {\bibfield
  {journal} {\bibinfo  {journal} {Phys. Rev. B}\ }\textbf {\bibinfo {volume}
  {23}},\ \bibinfo {pages} {5632} (\bibinfo {year} {1981})}\BibitemShut
  {NoStop}%
\bibitem [{\citenamefont {Halperin}(1982)}]{Halperin1982}%
  \BibitemOpen
  \bibfield  {author} {\bibinfo {author} {\bibfnamefont {B.~I.}\ \bibnamefont
  {Halperin}},\ }\href {\doibase 10.1103/PhysRevB.25.2185} {\bibfield
  {journal} {\bibinfo  {journal} {Phys. Rev. B}\ }\textbf {\bibinfo {volume}
  {25}},\ \bibinfo {pages} {2185} (\bibinfo {year} {1982})}\BibitemShut
  {NoStop}%
\bibitem [{\citenamefont {Wu}\ \emph {et~al.}(2006)\citenamefont {Wu},
  \citenamefont {Bernevig},\ and\ \citenamefont {Zhang}}]{Wu2006}%
  \BibitemOpen
  \bibfield  {author} {\bibinfo {author} {\bibfnamefont {C.}~\bibnamefont
  {Wu}}, \bibinfo {author} {\bibfnamefont {B.~A.}\ \bibnamefont {Bernevig}}, \
  and\ \bibinfo {author} {\bibfnamefont {S.~C.}\ \bibnamefont {Zhang}},\ }\href
  {\doibase 10.1103/PhysRevLett.96.106401} {\bibfield  {journal} {\bibinfo
  {journal} {Phys. Rev. Lett.}\ }\textbf {\bibinfo {volume} {96}},\ \bibinfo
  {pages} {106401} (\bibinfo {year} {2006})}\BibitemShut {NoStop}%
\bibitem [{\citenamefont {Bernevig}\ \emph {et~al.}(2006)\citenamefont
  {Bernevig}, \citenamefont {Hughes},\ and\ \citenamefont
  {Zhang}}]{Bernevig2006}%
  \BibitemOpen
  \bibfield  {author} {\bibinfo {author} {\bibfnamefont {B.~A.}\ \bibnamefont
  {Bernevig}}, \bibinfo {author} {\bibfnamefont {T.~L.}\ \bibnamefont
  {Hughes}}, \ and\ \bibinfo {author} {\bibfnamefont {S.-C.}\ \bibnamefont
  {Zhang}},\ }\href {\doibase 10.1126/science.1133734} {\bibfield  {journal}
  {\bibinfo  {journal} {Science (80-. ).}\ }\textbf {\bibinfo {volume} {314}},\
  \bibinfo {pages} {1757} (\bibinfo {year} {2006})}\BibitemShut {NoStop}%
\bibitem [{\citenamefont {Onoda}\ and\ \citenamefont
  {Nagaosa}(2005)}]{Onoda2005}%
  \BibitemOpen
  \bibfield  {author} {\bibinfo {author} {\bibfnamefont {M.}~\bibnamefont
  {Onoda}}\ and\ \bibinfo {author} {\bibfnamefont {N.}~\bibnamefont
  {Nagaosa}},\ }\href {\doibase 10.1103/PhysRevLett.95.106601} {\bibfield
  {journal} {\bibinfo  {journal} {Phys. Rev. Lett.}\ }\textbf {\bibinfo
  {volume} {95}},\ \bibinfo {pages} {106601} (\bibinfo {year}
  {2005})}\BibitemShut {NoStop}%
\bibitem [{\citenamefont {Thouless}\ \emph {et~al.}(1982)\citenamefont
  {Thouless}, \citenamefont {Kohmoto}, \citenamefont {Nightingale},\ and\
  \citenamefont {den Nijs}}]{Thouless1982}%
  \BibitemOpen
  \bibfield  {author} {\bibinfo {author} {\bibfnamefont {D.~J.}\ \bibnamefont
  {Thouless}}, \bibinfo {author} {\bibfnamefont {M.}~\bibnamefont {Kohmoto}},
  \bibinfo {author} {\bibfnamefont {M.~P.}\ \bibnamefont {Nightingale}}, \ and\
  \bibinfo {author} {\bibfnamefont {M.}~\bibnamefont {den Nijs}},\ }\href
  {\doibase 10.1103/PhysRevLett.49.405} {\bibfield  {journal} {\bibinfo
  {journal} {Phys. Rev. Lett.}\ }\textbf {\bibinfo {volume} {49}},\ \bibinfo
  {pages} {405} (\bibinfo {year} {1982})}\BibitemShut {NoStop}%
\bibitem [{\citenamefont {Hatsugai}(1993{\natexlab{a}})}]{Hatsugai1993}%
  \BibitemOpen
  \bibfield  {author} {\bibinfo {author} {\bibfnamefont {Y.}~\bibnamefont
  {Hatsugai}},\ }\href {\doibase 10.1103/PhysRevB.48.11851} {\bibfield
  {journal} {\bibinfo  {journal} {Phys. Rev. B}\ }\textbf {\bibinfo {volume}
  {48}},\ \bibinfo {pages} {11851} (\bibinfo {year}
  {1993}{\natexlab{a}})}\BibitemShut {NoStop}%
\bibitem [{\citenamefont {Hatsugai}(1993{\natexlab{b}})}]{Hatsugai1993a}%
  \BibitemOpen
  \bibfield  {author} {\bibinfo {author} {\bibfnamefont {Y.}~\bibnamefont
  {Hatsugai}},\ }\href {\doibase 10.1103/PhysRevLett.71.3697} {\bibfield
  {journal} {\bibinfo  {journal} {Phys. Rev. Lett.}\ }\textbf {\bibinfo
  {volume} {71}},\ \bibinfo {pages} {3697} (\bibinfo {year}
  {1993}{\natexlab{b}})}\BibitemShut {NoStop}%
\bibitem [{\citenamefont {Qi}\ \emph {et~al.}(2006)\citenamefont {Qi},
  \citenamefont {Wu},\ and\ \citenamefont {Zhang}}]{Qi2006}%
  \BibitemOpen
  \bibfield  {author} {\bibinfo {author} {\bibfnamefont {X.~L.}\ \bibnamefont
  {Qi}}, \bibinfo {author} {\bibfnamefont {Y.~S.}\ \bibnamefont {Wu}}, \ and\
  \bibinfo {author} {\bibfnamefont {S.~C.}\ \bibnamefont {Zhang}},\ }\href
  {\doibase 10.1103/PhysRevB.74.045125} {\bibfield  {journal} {\bibinfo
  {journal} {Phys. Rev. B - Condens. Matter Mater. Phys.}\ }\textbf {\bibinfo
  {volume} {74}},\ \bibinfo {pages} {045125} (\bibinfo {year}
  {2006})}\BibitemShut {NoStop}%
\bibitem [{\citenamefont {Onose}\ \emph {et~al.}(2010)\citenamefont {Onose},
  \citenamefont {Ideue}, \citenamefont {Katsura}, \citenamefont {Shiomi},
  \citenamefont {Nagaosa},\ and\ \citenamefont {Tokura}}]{Onose2010}%
  \BibitemOpen
  \bibfield  {author} {\bibinfo {author} {\bibfnamefont {Y.}~\bibnamefont
  {Onose}}, \bibinfo {author} {\bibfnamefont {T.}~\bibnamefont {Ideue}},
  \bibinfo {author} {\bibfnamefont {H.}~\bibnamefont {Katsura}}, \bibinfo
  {author} {\bibfnamefont {Y.}~\bibnamefont {Shiomi}}, \bibinfo {author}
  {\bibfnamefont {N.}~\bibnamefont {Nagaosa}}, \ and\ \bibinfo {author}
  {\bibfnamefont {Y.}~\bibnamefont {Tokura}},\ }\href {\doibase
  10.1126/science.1188260} {\bibfield  {journal} {\bibinfo  {journal}
  {Science}\ }\textbf {\bibinfo {volume} {329}},\ \bibinfo {pages} {297}
  (\bibinfo {year} {2010})}\BibitemShut {NoStop}%
\bibitem [{\citenamefont {Madon}\ \emph {et~al.}(2014)\citenamefont {Madon},
  \citenamefont {Pham}, \citenamefont {Lacour}, \citenamefont {Anane},
  \citenamefont {Cros}, \citenamefont {Hehn},\ and\ \citenamefont
  {Wegrowe}}]{Madon2014}%
  \BibitemOpen
  \bibfield  {author} {\bibinfo {author} {\bibfnamefont {B.}~\bibnamefont
  {Madon}}, \bibinfo {author} {\bibfnamefont {D.~C.}\ \bibnamefont {Pham}},
  \bibinfo {author} {\bibfnamefont {D.}~\bibnamefont {Lacour}}, \bibinfo
  {author} {\bibfnamefont {A.}~\bibnamefont {Anane}}, \bibinfo {author}
  {\bibfnamefont {R.~B.~V.}\ \bibnamefont {Cros}}, \bibinfo {author}
  {\bibfnamefont {M.}~\bibnamefont {Hehn}}, \ and\ \bibinfo {author}
  {\bibfnamefont {J.~E.}\ \bibnamefont {Wegrowe}},\ }\href
  {http://arxiv.org/abs/1412.3723} {\  (\bibinfo {year} {2014})},\ \Eprint
  {http://arxiv.org/abs/1412.3723} {arXiv:1412.3723} \BibitemShut {NoStop}%
\bibitem [{\citenamefont {Mochizuki}\ \emph {et~al.}(2014)\citenamefont
  {Mochizuki}, \citenamefont {Yu}, \citenamefont {Seki}, \citenamefont
  {Kanazawa}, \citenamefont {Koshibae}, \citenamefont {Zang}, \citenamefont
  {Mostovoy}, \citenamefont {Tokura},\ and\ \citenamefont
  {Nagaosa}}]{Mochizuki2014}%
  \BibitemOpen
  \bibfield  {author} {\bibinfo {author} {\bibfnamefont {M.}~\bibnamefont
  {Mochizuki}}, \bibinfo {author} {\bibfnamefont {X.~Z.}\ \bibnamefont {Yu}},
  \bibinfo {author} {\bibfnamefont {S.}~\bibnamefont {Seki}}, \bibinfo {author}
  {\bibfnamefont {N.}~\bibnamefont {Kanazawa}}, \bibinfo {author}
  {\bibfnamefont {W.}~\bibnamefont {Koshibae}}, \bibinfo {author}
  {\bibfnamefont {J.}~\bibnamefont {Zang}}, \bibinfo {author} {\bibfnamefont
  {M.}~\bibnamefont {Mostovoy}}, \bibinfo {author} {\bibfnamefont
  {Y.}~\bibnamefont {Tokura}}, \ and\ \bibinfo {author} {\bibfnamefont
  {N.}~\bibnamefont {Nagaosa}},\ }\href {\doibase 10.1038/nmat3862} {\bibfield
  {journal} {\bibinfo  {journal} {Nat. Mater.}\ }\textbf {\bibinfo {volume}
  {13}},\ \bibinfo {pages} {241} (\bibinfo {year} {2014})}\BibitemShut
  {NoStop}%
\bibitem [{\citenamefont {Cao}\ \emph {et~al.}(2015)\citenamefont {Cao},
  \citenamefont {Chen},\ and\ \citenamefont {He}}]{Cao2015}%
  \BibitemOpen
  \bibfield  {author} {\bibinfo {author} {\bibfnamefont {X.}~\bibnamefont
  {Cao}}, \bibinfo {author} {\bibfnamefont {K.}~\bibnamefont {Chen}}, \ and\
  \bibinfo {author} {\bibfnamefont {D.}~\bibnamefont {He}},\ }\href {\doibase
  10.1088/0953-8984/27/16/166003} {\bibfield  {journal} {\bibinfo  {journal}
  {J. Phys. Condens. Matter}\ }\textbf {\bibinfo {volume} {27}},\ \bibinfo
  {pages} {166003} (\bibinfo {year} {2015})}\BibitemShut {NoStop}%
\bibitem [{\citenamefont {Owerre}(2016{\natexlab{a}})}]{Owerre2016c}%
  \BibitemOpen
  \bibfield  {author} {\bibinfo {author} {\bibfnamefont {S.~A.}\ \bibnamefont
  {Owerre}},\ }\href {\doibase 10.1063/1.4959815} {\bibfield  {journal}
  {\bibinfo  {journal} {J. Appl. Phys.}\ }\textbf {\bibinfo {volume} {120}},\
  \bibinfo {pages} {43903} (\bibinfo {year} {2016}{\natexlab{a}})}\BibitemShut
  {NoStop}%
\bibitem [{\citenamefont {Tanabe}\ \emph {et~al.}(2016)\citenamefont {Tanabe},
  \citenamefont {Matsumoto}, \citenamefont {Ohe}, \citenamefont {Murakami},
  \citenamefont {Moriyama}, \citenamefont {Chiba}, \citenamefont {Kobayashi},\
  and\ \citenamefont {Ono}}]{Tanabe2016}%
  \BibitemOpen
  \bibfield  {author} {\bibinfo {author} {\bibfnamefont {K.}~\bibnamefont
  {Tanabe}}, \bibinfo {author} {\bibfnamefont {R.}~\bibnamefont {Matsumoto}},
  \bibinfo {author} {\bibfnamefont {J.-I.}\ \bibnamefont {Ohe}}, \bibinfo
  {author} {\bibfnamefont {S.}~\bibnamefont {Murakami}}, \bibinfo {author}
  {\bibfnamefont {T.}~\bibnamefont {Moriyama}}, \bibinfo {author}
  {\bibfnamefont {D.}~\bibnamefont {Chiba}}, \bibinfo {author} {\bibfnamefont
  {K.}~\bibnamefont {Kobayashi}}, \ and\ \bibinfo {author} {\bibfnamefont
  {T.}~\bibnamefont {Ono}},\ }\href {\doibase 10.1002/pssb.201552520}
  {\bibfield  {journal} {\bibinfo  {journal} {Phys. status solidi}\ }\textbf
  {\bibinfo {volume} {253}},\ \bibinfo {pages} {783} (\bibinfo {year}
  {2016})}\BibitemShut {NoStop}%
\bibitem [{\citenamefont {Ideue}\ \emph {et~al.}(2012)\citenamefont {Ideue},
  \citenamefont {Onose}, \citenamefont {Katsura}, \citenamefont {Shiomi},
  \citenamefont {Ishiwata}, \citenamefont {Nagaosa},\ and\ \citenamefont
  {Tokura}}]{Ideue2012}%
  \BibitemOpen
  \bibfield  {author} {\bibinfo {author} {\bibfnamefont {T.}~\bibnamefont
  {Ideue}}, \bibinfo {author} {\bibfnamefont {Y.}~\bibnamefont {Onose}},
  \bibinfo {author} {\bibfnamefont {H.}~\bibnamefont {Katsura}}, \bibinfo
  {author} {\bibfnamefont {Y.}~\bibnamefont {Shiomi}}, \bibinfo {author}
  {\bibfnamefont {S.}~\bibnamefont {Ishiwata}}, \bibinfo {author}
  {\bibfnamefont {N.}~\bibnamefont {Nagaosa}}, \ and\ \bibinfo {author}
  {\bibfnamefont {Y.}~\bibnamefont {Tokura}},\ }\href {\doibase
  10.1103/PhysRevB.85.134411} {\bibfield  {journal} {\bibinfo  {journal} {Phys.
  Rev. B}\ }\textbf {\bibinfo {volume} {85}},\ \bibinfo {pages} {134411}
  (\bibinfo {year} {2012})},\ \Eprint {http://arxiv.org/abs/1201.3002}
  {arXiv:1201.3002} \BibitemShut {NoStop}%
\bibitem [{\citenamefont {Hirschberger}\ \emph
  {et~al.}(2015{\natexlab{a}})\citenamefont {Hirschberger}, \citenamefont
  {Chisnell}, \citenamefont {Lee},\ and\ \citenamefont
  {Ong}}]{Hirschberger2015}%
  \BibitemOpen
  \bibfield  {author} {\bibinfo {author} {\bibfnamefont {M.}~\bibnamefont
  {Hirschberger}}, \bibinfo {author} {\bibfnamefont {R.}~\bibnamefont
  {Chisnell}}, \bibinfo {author} {\bibfnamefont {Y.~S.}\ \bibnamefont {Lee}}, \
  and\ \bibinfo {author} {\bibfnamefont {N.~P.}\ \bibnamefont {Ong}},\ }\href
  {\doibase 10.1103/PhysRevLett.115.106603} {\bibfield  {journal} {\bibinfo
  {journal} {Phys. Rev. Lett.}\ }\textbf {\bibinfo {volume} {115}},\ \bibinfo
  {pages} {106603} (\bibinfo {year} {2015}{\natexlab{a}})}\BibitemShut
  {NoStop}%
\bibitem [{\citenamefont {Hirschberger}\ \emph
  {et~al.}(2015{\natexlab{b}})\citenamefont {Hirschberger}, \citenamefont
  {Krizan}, \citenamefont {Cava},\ and\ \citenamefont
  {Ong}}]{Hirschberger2015a}%
  \BibitemOpen
  \bibfield  {author} {\bibinfo {author} {\bibfnamefont {M.}~\bibnamefont
  {Hirschberger}}, \bibinfo {author} {\bibfnamefont {J.~W.}\ \bibnamefont
  {Krizan}}, \bibinfo {author} {\bibfnamefont {R.~J.}\ \bibnamefont {Cava}}, \
  and\ \bibinfo {author} {\bibfnamefont {N.~P.}\ \bibnamefont {Ong}},\ }\href
  {\doibase 10.1126/science.1257340} {\bibfield  {journal} {\bibinfo  {journal}
  {Science (80-. ).}\ }\textbf {\bibinfo {volume} {348}},\ \bibinfo {pages}
  {106} (\bibinfo {year} {2015}{\natexlab{b}})}\BibitemShut {NoStop}%
\bibitem [{\citenamefont {Chisnell}\ \emph {et~al.}(2015)\citenamefont
  {Chisnell}, \citenamefont {Helton}, \citenamefont {Freedman}, \citenamefont
  {Singh}, \citenamefont {Bewley}, \citenamefont {Nocera},\ and\ \citenamefont
  {Lee}}]{Chisnell2015}%
  \BibitemOpen
  \bibfield  {author} {\bibinfo {author} {\bibfnamefont {R.}~\bibnamefont
  {Chisnell}}, \bibinfo {author} {\bibfnamefont {J.~S.}\ \bibnamefont
  {Helton}}, \bibinfo {author} {\bibfnamefont {D.~E.}\ \bibnamefont
  {Freedman}}, \bibinfo {author} {\bibfnamefont {D.~K.}\ \bibnamefont {Singh}},
  \bibinfo {author} {\bibfnamefont {R.~I.}\ \bibnamefont {Bewley}}, \bibinfo
  {author} {\bibfnamefont {D.~G.}\ \bibnamefont {Nocera}}, \ and\ \bibinfo
  {author} {\bibfnamefont {Y.~S.}\ \bibnamefont {Lee}},\ }\href {\doibase
  10.1103/PhysRevLett.115.147201} {\bibfield  {journal} {\bibinfo  {journal}
  {Phys. Rev. Lett.}\ }\textbf {\bibinfo {volume} {115}},\ \bibinfo {pages}
  {147201} (\bibinfo {year} {2015})}\BibitemShut {NoStop}%
\bibitem [{\citenamefont {Chen}\ \emph {et~al.}(2018)\citenamefont {Chen},
  \citenamefont {Chung}, \citenamefont {Gao}, \citenamefont {Chen},
  \citenamefont {Stone}, \citenamefont {Kolesnikov}, \citenamefont {Huang},\
  and\ \citenamefont {Dai}}]{Chen2018}%
  \BibitemOpen
  \bibfield  {author} {\bibinfo {author} {\bibfnamefont {L.}~\bibnamefont
  {Chen}}, \bibinfo {author} {\bibfnamefont {J.-H.}\ \bibnamefont {Chung}},
  \bibinfo {author} {\bibfnamefont {B.}~\bibnamefont {Gao}}, \bibinfo {author}
  {\bibfnamefont {T.}~\bibnamefont {Chen}}, \bibinfo {author} {\bibfnamefont
  {M.~B.}\ \bibnamefont {Stone}}, \bibinfo {author} {\bibfnamefont {A.~I.}\
  \bibnamefont {Kolesnikov}}, \bibinfo {author} {\bibfnamefont
  {Q.}~\bibnamefont {Huang}}, \ and\ \bibinfo {author} {\bibfnamefont
  {P.}~\bibnamefont {Dai}},\ }\href {\doibase 10.1103/PhysRevX.8.041028}
  {\bibfield  {journal} {\bibinfo  {journal} {Phys. Rev. X}\ }\textbf {\bibinfo
  {volume} {8}},\ \bibinfo {pages} {041028} (\bibinfo {year}
  {2018})}\BibitemShut {NoStop}%
\bibitem [{\citenamefont {Mook}\ \emph
  {et~al.}(2014{\natexlab{a}})\citenamefont {Mook}, \citenamefont {Henk},\ and\
  \citenamefont {Mertig}}]{Mook2014a}%
  \BibitemOpen
  \bibfield  {author} {\bibinfo {author} {\bibfnamefont {A.}~\bibnamefont
  {Mook}}, \bibinfo {author} {\bibfnamefont {J.}~\bibnamefont {Henk}}, \ and\
  \bibinfo {author} {\bibfnamefont {I.}~\bibnamefont {Mertig}},\ }\href
  {\doibase 10.1103/PhysRevB.89.134409} {\bibfield  {journal} {\bibinfo
  {journal} {Phys. Rev. B - Condens. Matter Mater. Phys.}\ }\textbf {\bibinfo
  {volume} {89}},\ \bibinfo {pages} {134409} (\bibinfo {year}
  {2014}{\natexlab{a}})}\BibitemShut {NoStop}%
\bibitem [{\citenamefont {Mook}\ \emph
  {et~al.}(2014{\natexlab{b}})\citenamefont {Mook}, \citenamefont {Henk},\ and\
  \citenamefont {Mertig}}]{Mook2014}%
  \BibitemOpen
  \bibfield  {author} {\bibinfo {author} {\bibfnamefont {A.}~\bibnamefont
  {Mook}}, \bibinfo {author} {\bibfnamefont {J.}~\bibnamefont {Henk}}, \ and\
  \bibinfo {author} {\bibfnamefont {I.}~\bibnamefont {Mertig}},\ }\href
  {\doibase 10.1103/PhysRevB.90.024412} {\bibfield  {journal} {\bibinfo
  {journal} {Phys. Rev. B}\ }\textbf {\bibinfo {volume} {90}},\ \bibinfo
  {pages} {024412} (\bibinfo {year} {2014}{\natexlab{b}})}\BibitemShut
  {NoStop}%
\bibitem [{\citenamefont {Seshadri}\ and\ \citenamefont
  {Sen}(2018)}]{Seshadri2018}%
  \BibitemOpen
  \bibfield  {author} {\bibinfo {author} {\bibfnamefont {R.}~\bibnamefont
  {Seshadri}}\ and\ \bibinfo {author} {\bibfnamefont {D.}~\bibnamefont {Sen}},\
  }\href {\doibase 10.1103/PhysRevB.97.134411} {\bibfield  {journal} {\bibinfo
  {journal} {Phys. Rev. B}\ }\textbf {\bibinfo {volume} {97}},\ \bibinfo
  {pages} {134411} (\bibinfo {year} {2018})}\BibitemShut {NoStop}%
\bibitem [{\citenamefont {Owerre}(2016{\natexlab{b}})}]{Owerre2016d}%
  \BibitemOpen
  \bibfield  {author} {\bibinfo {author} {\bibfnamefont {S.~A.}\ \bibnamefont
  {Owerre}},\ }\href {\doibase 10.1088/0953-8984/28/38/386001} {\bibfield
  {journal} {\bibinfo  {journal} {J. Phys. Condens. Matter}\ }\textbf {\bibinfo
  {volume} {28}},\ \bibinfo {pages} {386001} (\bibinfo {year}
  {2016}{\natexlab{b}})}\BibitemShut {NoStop}%
\bibitem [{\citenamefont {Kim}\ \emph {et~al.}(2016)\citenamefont {Kim},
  \citenamefont {Ochoa}, \citenamefont {Zarzuela},\ and\ \citenamefont
  {Tserkovnyak}}]{Kim2016a}%
  \BibitemOpen
  \bibfield  {author} {\bibinfo {author} {\bibfnamefont {S.~K.}\ \bibnamefont
  {Kim}}, \bibinfo {author} {\bibfnamefont {H.}~\bibnamefont {Ochoa}}, \bibinfo
  {author} {\bibfnamefont {R.}~\bibnamefont {Zarzuela}}, \ and\ \bibinfo
  {author} {\bibfnamefont {Y.}~\bibnamefont {Tserkovnyak}},\ }\href {\doibase
  10.1103/PhysRevLett.117.227201} {\bibfield  {journal} {\bibinfo  {journal}
  {Phys. Rev. Lett.}\ }\textbf {\bibinfo {volume} {117}},\ \bibinfo {pages}
  {227201} (\bibinfo {year} {2016})}\BibitemShut {NoStop}%
\bibitem [{\citenamefont {Amorim}\ \emph {et~al.}(2015)\citenamefont {Amorim},
  \citenamefont {Cortijo}, \citenamefont {{De Juan}}, \citenamefont {Grushin},
  \citenamefont {Guinea}, \citenamefont {Guti{\'{e}}rrez-Rubio}, \citenamefont
  {Ochoa}, \citenamefont {Parente}, \citenamefont {Rold{\'{a}}n}, \citenamefont
  {San-Jose}, \citenamefont {Schiefele}, \citenamefont {Sturla},\ and\
  \citenamefont {Vozmediano}}]{Amorim2016}%
  \BibitemOpen
  \bibfield  {author} {\bibinfo {author} {\bibfnamefont {B.}~\bibnamefont
  {Amorim}}, \bibinfo {author} {\bibfnamefont {A.}~\bibnamefont {Cortijo}},
  \bibinfo {author} {\bibfnamefont {F.}~\bibnamefont {{De Juan}}}, \bibinfo
  {author} {\bibfnamefont {A.~G.}\ \bibnamefont {Grushin}}, \bibinfo {author}
  {\bibfnamefont {F.}~\bibnamefont {Guinea}}, \bibinfo {author} {\bibfnamefont
  {A.}~\bibnamefont {Guti{\'{e}}rrez-Rubio}}, \bibinfo {author} {\bibfnamefont
  {H.}~\bibnamefont {Ochoa}}, \bibinfo {author} {\bibfnamefont
  {V.}~\bibnamefont {Parente}}, \bibinfo {author} {\bibfnamefont
  {R.}~\bibnamefont {Rold{\'{a}}n}}, \bibinfo {author} {\bibfnamefont
  {P.}~\bibnamefont {San-Jose}}, \bibinfo {author} {\bibfnamefont
  {J.}~\bibnamefont {Schiefele}}, \bibinfo {author} {\bibfnamefont
  {M.}~\bibnamefont {Sturla}}, \ and\ \bibinfo {author} {\bibfnamefont {M.~A.}\
  \bibnamefont {Vozmediano}},\ }\href {\doibase 10.1016/j.physrep.2015.12.006}
  {\bibfield  {journal} {\bibinfo  {journal} {Phys. Rep.}\ }\textbf {\bibinfo
  {volume} {617}},\ \bibinfo {pages} {1} (\bibinfo {year} {2015})}\BibitemShut
  {NoStop}%
\bibitem [{\citenamefont {Naumis}\ \emph {et~al.}(2017)\citenamefont {Naumis},
  \citenamefont {Barraza-Lopez}, \citenamefont {Oliva-Leyva},\ and\
  \citenamefont {Terrones}}]{Naumis2017}%
  \BibitemOpen
  \bibfield  {author} {\bibinfo {author} {\bibfnamefont {G.~G.}\ \bibnamefont
  {Naumis}}, \bibinfo {author} {\bibfnamefont {S.}~\bibnamefont
  {Barraza-Lopez}}, \bibinfo {author} {\bibfnamefont {M.}~\bibnamefont
  {Oliva-Leyva}}, \ and\ \bibinfo {author} {\bibfnamefont {H.}~\bibnamefont
  {Terrones}},\ }\href {\doibase 10.1088/1361-6633/aa74ef} {\bibfield
  {journal} {\bibinfo  {journal} {Reports Prog. Phys.}\ }\textbf {\bibinfo
  {volume} {80}},\ \bibinfo {pages} {096501} (\bibinfo {year}
  {2017})}\BibitemShut {NoStop}%
\bibitem [{\citenamefont {Gonz{\'{a}}lez-{\'{A}}rraga}\ \emph
  {et~al.}(2018)\citenamefont {Gonz{\'{a}}lez-{\'{A}}rraga}, \citenamefont
  {Guinea},\ and\ \citenamefont {San-Jose}}]{Gonzalez-Arraga2018}%
  \BibitemOpen
  \bibfield  {author} {\bibinfo {author} {\bibfnamefont {L.}~\bibnamefont
  {Gonz{\'{a}}lez-{\'{A}}rraga}}, \bibinfo {author} {\bibfnamefont
  {F.}~\bibnamefont {Guinea}}, \ and\ \bibinfo {author} {\bibfnamefont
  {P.}~\bibnamefont {San-Jose}},\ }\href {\doibase 10.1103/PhysRevB.97.165430}
  {\bibfield  {journal} {\bibinfo  {journal} {Phys. Rev. B}\ }\textbf {\bibinfo
  {volume} {97}},\ \bibinfo {pages} {165430} (\bibinfo {year}
  {2018})}\BibitemShut {NoStop}%
\bibitem [{\citenamefont {Kariyado}\ and\ \citenamefont
  {Hu}(2017)}]{Kariyado2017}%
  \BibitemOpen
  \bibfield  {author} {\bibinfo {author} {\bibfnamefont {T.}~\bibnamefont
  {Kariyado}}\ and\ \bibinfo {author} {\bibfnamefont {X.}~\bibnamefont {Hu}},\
  }\href {\doibase 10.1038/s41598-017-16334-0} {\bibfield  {journal} {\bibinfo
  {journal} {Sci. Rep.}\ }\textbf {\bibinfo {volume} {7}},\ \bibinfo {pages}
  {16515} (\bibinfo {year} {2017})}\BibitemShut {NoStop}%
\bibitem [{\citenamefont {Liu}\ \emph {et~al.}(2017)\citenamefont {Liu},
  \citenamefont {Yamamoto},\ and\ \citenamefont {Wakabayashi}}]{Liu2017}%
  \BibitemOpen
  \bibfield  {author} {\bibinfo {author} {\bibfnamefont {F.}~\bibnamefont
  {Liu}}, \bibinfo {author} {\bibfnamefont {M.}~\bibnamefont {Yamamoto}}, \
  and\ \bibinfo {author} {\bibfnamefont {K.}~\bibnamefont {Wakabayashi}},\
  }\href {\doibase 10.7566/JPSJ.86.123707} {\bibfield  {journal} {\bibinfo
  {journal} {J. Phys. Soc. Japan}\ }\textbf {\bibinfo {volume} {86}},\ \bibinfo
  {pages} {123707} (\bibinfo {year} {2017})}\BibitemShut {NoStop}%
\bibitem [{\citenamefont {Grandi}\ \emph {et~al.}(2015)\citenamefont {Grandi},
  \citenamefont {Manghi}, \citenamefont {Corradini},\ and\ \citenamefont
  {Bertoni}}]{Grandi2015}%
  \BibitemOpen
  \bibfield  {author} {\bibinfo {author} {\bibfnamefont {F.}~\bibnamefont
  {Grandi}}, \bibinfo {author} {\bibfnamefont {F.}~\bibnamefont {Manghi}},
  \bibinfo {author} {\bibfnamefont {O.}~\bibnamefont {Corradini}}, \ and\
  \bibinfo {author} {\bibfnamefont {C.~M.}\ \bibnamefont {Bertoni}},\ }\href
  {\doibase 10.1103/PhysRevB.91.115112} {\bibfield  {journal} {\bibinfo
  {journal} {Phys. Rev. B}\ }\textbf {\bibinfo {volume} {91}},\ \bibinfo
  {pages} {115112} (\bibinfo {year} {2015})}\BibitemShut {NoStop}%
\bibitem [{\citenamefont {Wu}\ and\ \citenamefont {Hu}(2016)}]{Wu2016}%
  \BibitemOpen
  \bibfield  {author} {\bibinfo {author} {\bibfnamefont {L.-H.}\ \bibnamefont
  {Wu}}\ and\ \bibinfo {author} {\bibfnamefont {X.}~\bibnamefont {Hu}},\ }\href
  {\doibase 10.1038/srep24347} {\bibfield  {journal} {\bibinfo  {journal} {Sci.
  Rep.}\ }\textbf {\bibinfo {volume} {6}},\ \bibinfo {pages} {24347} (\bibinfo
  {year} {2016})}\BibitemShut {NoStop}%
\bibitem [{\citenamefont {Ferreiros}\ and\ \citenamefont
  {Vozmediano}(2018)}]{Ferreiros2018}%
  \BibitemOpen
  \bibfield  {author} {\bibinfo {author} {\bibfnamefont {Y.}~\bibnamefont
  {Ferreiros}}\ and\ \bibinfo {author} {\bibfnamefont {M.~A.}\ \bibnamefont
  {Vozmediano}},\ }\href {\doibase 10.1103/PhysRevB.97.054404} {\bibfield
  {journal} {\bibinfo  {journal} {Phys. Rev. B}\ }\textbf {\bibinfo {volume}
  {97}},\ \bibinfo {pages} {054404} (\bibinfo {year} {2018})}\BibitemShut
  {NoStop}%
\bibitem [{\citenamefont {Owerre}(2018)}]{Owerre2018}%
  \BibitemOpen
  \bibfield  {author} {\bibinfo {author} {\bibfnamefont {S.~A.}\ \bibnamefont
  {Owerre}},\ }\href {\doibase 10.1088/1361-648X/aac365} {\bibfield  {journal}
  {\bibinfo  {journal} {J. Phys. Condens. Matter}\ }\textbf {\bibinfo {volume}
  {30}},\ \bibinfo {pages} {245803} (\bibinfo {year} {2018})},\ \Eprint
  {http://arxiv.org/abs/1802.10095} {arXiv:1802.10095} \BibitemShut {NoStop}%
\bibitem [{\citenamefont {Plotnik}\ \emph {et~al.}(2014)\citenamefont
  {Plotnik}, \citenamefont {Rechtsman}, \citenamefont {Song}, \citenamefont
  {Heinrich}, \citenamefont {Zeuner}, \citenamefont {Nolte}, \citenamefont
  {Lumer}, \citenamefont {Malkova}, \citenamefont {Xu}, \citenamefont
  {Szameit}, \citenamefont {Chen},\ and\ \citenamefont {Segev}}]{Plotnik2014}%
  \BibitemOpen
  \bibfield  {author} {\bibinfo {author} {\bibfnamefont {Y.}~\bibnamefont
  {Plotnik}}, \bibinfo {author} {\bibfnamefont {M.~C.}\ \bibnamefont
  {Rechtsman}}, \bibinfo {author} {\bibfnamefont {D.}~\bibnamefont {Song}},
  \bibinfo {author} {\bibfnamefont {M.}~\bibnamefont {Heinrich}}, \bibinfo
  {author} {\bibfnamefont {J.~M.}\ \bibnamefont {Zeuner}}, \bibinfo {author}
  {\bibfnamefont {S.}~\bibnamefont {Nolte}}, \bibinfo {author} {\bibfnamefont
  {Y.}~\bibnamefont {Lumer}}, \bibinfo {author} {\bibfnamefont
  {N.}~\bibnamefont {Malkova}}, \bibinfo {author} {\bibfnamefont
  {J.}~\bibnamefont {Xu}}, \bibinfo {author} {\bibfnamefont {A.}~\bibnamefont
  {Szameit}}, \bibinfo {author} {\bibfnamefont {Z.}~\bibnamefont {Chen}}, \
  and\ \bibinfo {author} {\bibfnamefont {M.}~\bibnamefont {Segev}},\ }\href
  {\doibase 10.1038/nmat3783} {\bibfield  {journal} {\bibinfo  {journal} {Nat.
  Mater.}\ }\textbf {\bibinfo {volume} {13}},\ \bibinfo {pages} {57} (\bibinfo
  {year} {2014})}\BibitemShut {NoStop}%
\bibitem [{\citenamefont {Pantale{\'{o}}n}\ and\ \citenamefont
  {Xian}(2018{\natexlab{a}})}]{Pantaleon2018a}%
  \BibitemOpen
  \bibfield  {author} {\bibinfo {author} {\bibfnamefont {P.~A.}\ \bibnamefont
  {Pantale{\'{o}}n}}\ and\ \bibinfo {author} {\bibfnamefont {Y.}~\bibnamefont
  {Xian}},\ }\href {\doibase 10.7566/JPSJ.87.064005} {\bibfield  {journal}
  {\bibinfo  {journal} {J. Phys. Soc. Japan}\ }\textbf {\bibinfo {volume}
  {87}},\ \bibinfo {pages} {064005} (\bibinfo {year}
  {2018}{\natexlab{a}})}\BibitemShut {NoStop}%
\bibitem [{\citenamefont {Moriya}(1960)}]{Moriya1960}%
  \BibitemOpen
  \bibfield  {author} {\bibinfo {author} {\bibfnamefont {T.}~\bibnamefont
  {Moriya}},\ }\href {\doibase 10.1103/PhysRev.120.91} {\bibfield  {journal}
  {\bibinfo  {journal} {Phys. Rev.}\ }\textbf {\bibinfo {volume} {120}},\
  \bibinfo {pages} {91} (\bibinfo {year} {1960})}\BibitemShut {NoStop}%
\bibitem [{\citenamefont {Wu}\ \emph {et~al.}(2012)\citenamefont {Wu},
  \citenamefont {Rachel}, \citenamefont {Liu},\ and\ \citenamefont {{Le
  Hur}}}]{Wu2012}%
  \BibitemOpen
  \bibfield  {author} {\bibinfo {author} {\bibfnamefont {W.}~\bibnamefont
  {Wu}}, \bibinfo {author} {\bibfnamefont {S.}~\bibnamefont {Rachel}}, \bibinfo
  {author} {\bibfnamefont {W.-M.}\ \bibnamefont {Liu}}, \ and\ \bibinfo
  {author} {\bibfnamefont {K.}~\bibnamefont {{Le Hur}}},\ }\href {\doibase
  10.1103/PhysRevB.85.205102} {\bibfield  {journal} {\bibinfo  {journal} {Phys.
  Rev. B}\ }\textbf {\bibinfo {volume} {85}},\ \bibinfo {pages} {205102}
  (\bibinfo {year} {2012})}\BibitemShut {NoStop}%
\bibitem [{\citenamefont {Akhiezer}\ \emph {et~al.}(1961)\citenamefont
  {Akhiezer}, \citenamefont {Bar'yakhtar},\ and\ \citenamefont
  {Kaganov}}]{Akhiezer1961}%
  \BibitemOpen
  \bibfield  {author} {\bibinfo {author} {\bibfnamefont {A.~I.}\ \bibnamefont
  {Akhiezer}}, \bibinfo {author} {\bibfnamefont {V.~G.}\ \bibnamefont
  {Bar'yakhtar}}, \ and\ \bibinfo {author} {\bibfnamefont {M.~I.}\ \bibnamefont
  {Kaganov}},\ }\href {\doibase 10.1070/PU1961v003n04ABEH003309} {\bibfield
  {journal} {\bibinfo  {journal} {Sov. Phys. Uspekhi}\ }\textbf {\bibinfo
  {volume} {3}},\ \bibinfo {pages} {567} (\bibinfo {year} {1961})}\BibitemShut
  {NoStop}%
\bibitem [{\citenamefont {Chamon}(2000)}]{Chamon2000}%
  \BibitemOpen
  \bibfield  {author} {\bibinfo {author} {\bibfnamefont {C.}~\bibnamefont
  {Chamon}},\ }\href {\doibase 10.1103/PhysRevB.62.2806} {\bibfield  {journal}
  {\bibinfo  {journal} {Phys. Rev. B}\ }\textbf {\bibinfo {volume} {62}},\
  \bibinfo {pages} {2806} (\bibinfo {year} {2000})}\BibitemShut {NoStop}%
\bibitem [{\citenamefont {Hou}\ \emph {et~al.}(2007)\citenamefont {Hou},
  \citenamefont {Chamon},\ and\ \citenamefont {Mudry}}]{Hou2007}%
  \BibitemOpen
  \bibfield  {author} {\bibinfo {author} {\bibfnamefont {C.-Y.}\ \bibnamefont
  {Hou}}, \bibinfo {author} {\bibfnamefont {C.}~\bibnamefont {Chamon}}, \ and\
  \bibinfo {author} {\bibfnamefont {C.}~\bibnamefont {Mudry}},\ }\href
  {\doibase 10.1103/PhysRevLett.98.186809} {\bibfield  {journal} {\bibinfo
  {journal} {Phys. Rev. Lett.}\ }\textbf {\bibinfo {volume} {98}},\ \bibinfo
  {pages} {186809} (\bibinfo {year} {2007})}\BibitemShut {NoStop}%
\bibitem [{\citenamefont {Gamayun}\ \emph {et~al.}(2018)\citenamefont
  {Gamayun}, \citenamefont {Ostroukh}, \citenamefont {Gnezdilov}, \citenamefont
  {Adagideli},\ and\ \citenamefont {Beenakker}}]{Gamayun2018}%
  \BibitemOpen
  \bibfield  {author} {\bibinfo {author} {\bibfnamefont {O.~V.}\ \bibnamefont
  {Gamayun}}, \bibinfo {author} {\bibfnamefont {V.~P.}\ \bibnamefont
  {Ostroukh}}, \bibinfo {author} {\bibfnamefont {N.~V.}\ \bibnamefont
  {Gnezdilov}}, \bibinfo {author} {\bibfnamefont {Ä.}~\bibnamefont
  {Adagideli}}, \ and\ \bibinfo {author} {\bibfnamefont {C.~W.~J.}\
  \bibnamefont {Beenakker}},\ }\href {\doibase 10.1088/1367-2630/aaa7e5}
  {\bibfield  {journal} {\bibinfo  {journal} {New J. Phys.}\ }\textbf {\bibinfo
  {volume} {20}},\ \bibinfo {pages} {023016} (\bibinfo {year}
  {2018})}\BibitemShut {NoStop}%
\bibitem [{\citenamefont {Pershoguba}\ \emph {et~al.}(2018)\citenamefont
  {Pershoguba}, \citenamefont {Banerjee}, \citenamefont {Lashley},
  \citenamefont {Park}, \citenamefont {{\AA}gren}, \citenamefont {Aeppli},\
  and\ \citenamefont {Balatsky}}]{Pershoguba2017}%
  \BibitemOpen
  \bibfield  {author} {\bibinfo {author} {\bibfnamefont {S.~S.}\ \bibnamefont
  {Pershoguba}}, \bibinfo {author} {\bibfnamefont {S.}~\bibnamefont
  {Banerjee}}, \bibinfo {author} {\bibfnamefont {J.~C.~C.}\ \bibnamefont
  {Lashley}}, \bibinfo {author} {\bibfnamefont {J.}~\bibnamefont {Park}},
  \bibinfo {author} {\bibfnamefont {H.}~\bibnamefont {{\AA}gren}}, \bibinfo
  {author} {\bibfnamefont {G.}~\bibnamefont {Aeppli}}, \ and\ \bibinfo {author}
  {\bibfnamefont {A.~V.}\ \bibnamefont {Balatsky}},\ }\href {\doibase
  10.1103/PhysRevX.8.011010} {\bibfield  {journal} {\bibinfo  {journal} {Phys.
  Rev. X}\ }\textbf {\bibinfo {volume} {8}},\ \bibinfo {pages} {011010}
  (\bibinfo {year} {2018})}\BibitemShut {NoStop}%
\bibitem [{\citenamefont {Murakami}\ \emph {et~al.}(2017)\citenamefont
  {Murakami}, \citenamefont {Hirayama}, \citenamefont {Okugawa},\ and\
  \citenamefont {Miyake}}]{Murakami2017}%
  \BibitemOpen
  \bibfield  {author} {\bibinfo {author} {\bibfnamefont {S.}~\bibnamefont
  {Murakami}}, \bibinfo {author} {\bibfnamefont {M.}~\bibnamefont {Hirayama}},
  \bibinfo {author} {\bibfnamefont {R.}~\bibnamefont {Okugawa}}, \ and\
  \bibinfo {author} {\bibfnamefont {T.}~\bibnamefont {Miyake}},\ }\href
  {\doibase 10.1126/sciadv.1602680} {\bibfield  {journal} {\bibinfo  {journal}
  {Sci. Adv.}\ }\textbf {\bibinfo {volume} {3}},\ \bibinfo {pages} {e1602680}
  (\bibinfo {year} {2017})}\BibitemShut {NoStop}%
\bibitem [{\citenamefont {Fransson}\ \emph {et~al.}(2016)\citenamefont
  {Fransson}, \citenamefont {Black-Schaffer},\ and\ \citenamefont
  {Balatsky}}]{Fransson2016}%
  \BibitemOpen
  \bibfield  {author} {\bibinfo {author} {\bibfnamefont {J.}~\bibnamefont
  {Fransson}}, \bibinfo {author} {\bibfnamefont {A.~M.}\ \bibnamefont
  {Black-Schaffer}}, \ and\ \bibinfo {author} {\bibfnamefont {A.~V.}\
  \bibnamefont {Balatsky}},\ }\href {\doibase 10.1103/PhysRevB.94.075401}
  {\bibfield  {journal} {\bibinfo  {journal} {Phys. Rev. B}\ }\textbf {\bibinfo
  {volume} {94}},\ \bibinfo {pages} {75401} (\bibinfo {year}
  {2016})}\BibitemShut {NoStop}%
\bibitem [{\citenamefont {Berry}(1984)}]{berry84}%
  \BibitemOpen
  \bibfield  {author} {\bibinfo {author} {\bibfnamefont {M.~V.}\ \bibnamefont
  {Berry}},\ }\href {http://www.jstor.org/stable/2397741} {\bibfield  {journal}
  {\bibinfo  {journal} {Proc. R. Soc. Lond. A. Math. Phys. Sci.}\ }\textbf
  {\bibinfo {volume} {392}},\ \bibinfo {pages} {45} (\bibinfo {year}
  {1984})}\BibitemShut {NoStop}%
\bibitem [{\citenamefont {Fukui}\ \emph {et~al.}(2005)\citenamefont {Fukui},
  \citenamefont {Hatsugai},\ and\ \citenamefont {Suzuki}}]{Fukui2005}%
  \BibitemOpen
  \bibfield  {author} {\bibinfo {author} {\bibfnamefont {T.}~\bibnamefont
  {Fukui}}, \bibinfo {author} {\bibfnamefont {Y.}~\bibnamefont {Hatsugai}}, \
  and\ \bibinfo {author} {\bibfnamefont {H.}~\bibnamefont {Suzuki}},\ }\href
  {\doibase 10.1143/JPSJ.74.1674} {\bibfield  {journal} {\bibinfo  {journal}
  {J. Phys. Soc. Japan}\ }\textbf {\bibinfo {volume} {74}},\ \bibinfo {pages}
  {1674} (\bibinfo {year} {2005})}\BibitemShut {NoStop}%
\bibitem [{\citenamefont {Shindou}\ \emph {et~al.}(2013)\citenamefont
  {Shindou}, \citenamefont {Matsumoto}, \citenamefont {Murakami},\ and\
  \citenamefont {Ohe}}]{Shindou2013}%
  \BibitemOpen
  \bibfield  {author} {\bibinfo {author} {\bibfnamefont {R.}~\bibnamefont
  {Shindou}}, \bibinfo {author} {\bibfnamefont {R.}~\bibnamefont {Matsumoto}},
  \bibinfo {author} {\bibfnamefont {S.}~\bibnamefont {Murakami}}, \ and\
  \bibinfo {author} {\bibfnamefont {J.-i.}\ \bibnamefont {Ohe}},\ }\href
  {\doibase 10.1103/PhysRevB.87.174427} {\bibfield  {journal} {\bibinfo
  {journal} {Phys. Rev. B}\ }\textbf {\bibinfo {volume} {87}},\ \bibinfo
  {pages} {174427} (\bibinfo {year} {2013})}\BibitemShut {NoStop}%
\bibitem [{\citenamefont {{\'{C}}eli{\'{c}}}\ \emph {et~al.}(1996)\citenamefont
  {{\'{C}}eli{\'{c}}}, \citenamefont {Kapor}, \citenamefont {{\v{S}}krinjar},\
  and\ \citenamefont {Stojanovi{\'{c}}}}]{Celic1996}%
  \BibitemOpen
  \bibfield  {author} {\bibinfo {author} {\bibfnamefont {A.}~\bibnamefont
  {{\'{C}}eli{\'{c}}}}, \bibinfo {author} {\bibfnamefont {D.}~\bibnamefont
  {Kapor}}, \bibinfo {author} {\bibfnamefont {M.}~\bibnamefont
  {{\v{S}}krinjar}}, \ and\ \bibinfo {author} {\bibfnamefont {S.}~\bibnamefont
  {Stojanovi{\'{c}}}},\ }\href {\doibase 10.1016/0375-9601(96)00353-2}
  {\bibfield  {journal} {\bibinfo  {journal} {Phys. Lett. A}\ }\textbf
  {\bibinfo {volume} {219}},\ \bibinfo {pages} {121} (\bibinfo {year}
  {1996})}\BibitemShut {NoStop}%
\bibitem [{\citenamefont {You}\ \emph {et~al.}(2008)\citenamefont {You},
  \citenamefont {Huang},\ and\ \citenamefont {Lin}}]{You2008a}%
  \BibitemOpen
  \bibfield  {author} {\bibinfo {author} {\bibfnamefont {J.-S.}\ \bibnamefont
  {You}}, \bibinfo {author} {\bibfnamefont {W.-M.}\ \bibnamefont {Huang}}, \
  and\ \bibinfo {author} {\bibfnamefont {H.-H.}\ \bibnamefont {Lin}},\ }\href
  {\doibase 10.1103/PhysRevB.78.161404} {\bibfield  {journal} {\bibinfo
  {journal} {Phys. Rev. B}\ }\textbf {\bibinfo {volume} {78}},\ \bibinfo
  {pages} {161404} (\bibinfo {year} {2008})}\BibitemShut {NoStop}%
\bibitem [{\citenamefont {Sakaguchi}\ and\ \citenamefont
  {Matsumoto}(2016)}]{Sakaguchi2016}%
  \BibitemOpen
  \bibfield  {author} {\bibinfo {author} {\bibfnamefont {R.}~\bibnamefont
  {Sakaguchi}}\ and\ \bibinfo {author} {\bibfnamefont {M.}~\bibnamefont
  {Matsumoto}},\ }\href {\doibase 10.7566/JPSJ.85.104707} {\bibfield  {journal}
  {\bibinfo  {journal} {J. Phys. Soc. Japan}\ }\textbf {\bibinfo {volume}
  {85}},\ \bibinfo {pages} {104707} (\bibinfo {year} {2016})}\BibitemShut
  {NoStop}%
\bibitem [{\citenamefont {Ezawa}\ and\ \citenamefont
  {Nagaosa}(2013)}]{Ezawa2013}%
  \BibitemOpen
  \bibfield  {author} {\bibinfo {author} {\bibfnamefont {M.}~\bibnamefont
  {Ezawa}}\ and\ \bibinfo {author} {\bibfnamefont {N.}~\bibnamefont
  {Nagaosa}},\ }\href {\doibase 10.1103/PhysRevB.88.121401} {\bibfield
  {journal} {\bibinfo  {journal} {Phys. Rev. B}\ }\textbf {\bibinfo {volume}
  {88}},\ \bibinfo {pages} {121401} (\bibinfo {year} {2013})}\BibitemShut
  {NoStop}%
\bibitem [{\citenamefont {Pantale{\'{o}}n}\ and\ \citenamefont
  {Xian}(2018{\natexlab{b}})}]{Pantaleon2017a}%
  \BibitemOpen
  \bibfield  {author} {\bibinfo {author} {\bibfnamefont {P.~A.}\ \bibnamefont
  {Pantale{\'{o}}n}}\ and\ \bibinfo {author} {\bibfnamefont {Y.}~\bibnamefont
  {Xian}},\ }\href {\doibase 10.1016/j.physb.2017.11.040} {\bibfield  {journal}
  {\bibinfo  {journal} {Phys. B Condens. Matter}\ }\textbf {\bibinfo {volume}
  {530}},\ \bibinfo {pages} {191} (\bibinfo {year}
  {2018}{\natexlab{b}})}\BibitemShut {NoStop}%
\bibitem [{\citenamefont {Tamm}(1932)}]{Tamm}%
  \BibitemOpen
  \bibfield  {author} {\bibinfo {author} {\bibfnamefont {I.}~\bibnamefont
  {Tamm}},\ }\href@noop {} {\bibfield  {journal} {\bibinfo  {journal} {Phys. Z.
  Sov. Union}\ }\textbf {\bibinfo {volume} {1}},\ \bibinfo {pages} {733}
  (\bibinfo {year} {1932})}\BibitemShut {NoStop}%
\bibitem [{\citenamefont {Mook}\ \emph {et~al.}(2015)\citenamefont {Mook},
  \citenamefont {Henk},\ and\ \citenamefont {Mertig}}]{Mook2015}%
  \BibitemOpen
  \bibfield  {author} {\bibinfo {author} {\bibfnamefont {A.}~\bibnamefont
  {Mook}}, \bibinfo {author} {\bibfnamefont {J.}~\bibnamefont {Henk}}, \ and\
  \bibinfo {author} {\bibfnamefont {I.}~\bibnamefont {Mertig}},\ }\href
  {\doibase 10.1103/PhysRevB.91.174409} {\bibfield  {journal} {\bibinfo
  {journal} {Phys. Rev. B}\ }\textbf {\bibinfo {volume} {91}},\ \bibinfo
  {pages} {174409} (\bibinfo {year} {2015})}\BibitemShut {NoStop}%
\bibitem [{\citenamefont {Cheianov}\ \emph {et~al.}(2009)\citenamefont
  {Cheianov}, \citenamefont {Fal'ko}, \citenamefont {Sylju{\aa}sen},\ and\
  \citenamefont {Altshuler}}]{Cheianov2009a}%
  \BibitemOpen
  \bibfield  {author} {\bibinfo {author} {\bibfnamefont {V.}~\bibnamefont
  {Cheianov}}, \bibinfo {author} {\bibfnamefont {V.}~\bibnamefont {Fal'ko}},
  \bibinfo {author} {\bibfnamefont {O.}~\bibnamefont {Sylju{\aa}sen}}, \ and\
  \bibinfo {author} {\bibfnamefont {B.}~\bibnamefont {Altshuler}},\ }\href
  {\doibase 10.1016/j.ssc.2009.07.008} {\bibfield  {journal} {\bibinfo
  {journal} {Solid State Commun.}\ }\textbf {\bibinfo {volume} {149}},\
  \bibinfo {pages} {1499} (\bibinfo {year} {2009})},\ \Eprint
  {http://arxiv.org/abs/0906.5174} {arXiv:0906.5174} \BibitemShut {NoStop}%
\bibitem [{\citenamefont {Guti{\'{e}}rrez}\ \emph {et~al.}(2016)\citenamefont
  {Guti{\'{e}}rrez}, \citenamefont {Kim}, \citenamefont {Brown}, \citenamefont
  {Schiros}, \citenamefont {Nordlund}, \citenamefont {Lochocki}, \citenamefont
  {Shen}, \citenamefont {Park},\ and\ \citenamefont
  {Pasupathy}}]{Gutierrez2016}%
  \BibitemOpen
  \bibfield  {author} {\bibinfo {author} {\bibfnamefont {C.}~\bibnamefont
  {Guti{\'{e}}rrez}}, \bibinfo {author} {\bibfnamefont {C.-J.}\ \bibnamefont
  {Kim}}, \bibinfo {author} {\bibfnamefont {L.}~\bibnamefont {Brown}}, \bibinfo
  {author} {\bibfnamefont {T.}~\bibnamefont {Schiros}}, \bibinfo {author}
  {\bibfnamefont {D.}~\bibnamefont {Nordlund}}, \bibinfo {author}
  {\bibfnamefont {E.~B.}\ \bibnamefont {Lochocki}}, \bibinfo {author}
  {\bibfnamefont {K.~M.}\ \bibnamefont {Shen}}, \bibinfo {author}
  {\bibfnamefont {J.}~\bibnamefont {Park}}, \ and\ \bibinfo {author}
  {\bibfnamefont {A.~N.}\ \bibnamefont {Pasupathy}},\ }\href {\doibase
  10.1038/nphys3776} {\bibfield  {journal} {\bibinfo  {journal} {Nat. Phys.}\
  }\textbf {\bibinfo {volume} {12}},\ \bibinfo {pages} {950} (\bibinfo {year}
  {2016})}\BibitemShut {NoStop}%
\bibitem [{\citenamefont {Gong}\ \emph {et~al.}(2017)\citenamefont {Gong},
  \citenamefont {Li}, \citenamefont {Li}, \citenamefont {Ji}, \citenamefont
  {Stern}, \citenamefont {Xia}, \citenamefont {Cao}, \citenamefont {Bao},
  \citenamefont {Wang}, \citenamefont {Wang}, \citenamefont {Qiu},
  \citenamefont {Cava}, \citenamefont {Louie}, \citenamefont {Xia},\ and\
  \citenamefont {Zhang}}]{Gong2017}%
  \BibitemOpen
  \bibfield  {author} {\bibinfo {author} {\bibfnamefont {C.}~\bibnamefont
  {Gong}}, \bibinfo {author} {\bibfnamefont {L.}~\bibnamefont {Li}}, \bibinfo
  {author} {\bibfnamefont {Z.}~\bibnamefont {Li}}, \bibinfo {author}
  {\bibfnamefont {H.}~\bibnamefont {Ji}}, \bibinfo {author} {\bibfnamefont
  {A.}~\bibnamefont {Stern}}, \bibinfo {author} {\bibfnamefont
  {Y.}~\bibnamefont {Xia}}, \bibinfo {author} {\bibfnamefont {T.}~\bibnamefont
  {Cao}}, \bibinfo {author} {\bibfnamefont {W.}~\bibnamefont {Bao}}, \bibinfo
  {author} {\bibfnamefont {C.}~\bibnamefont {Wang}}, \bibinfo {author}
  {\bibfnamefont {Y.}~\bibnamefont {Wang}}, \bibinfo {author} {\bibfnamefont
  {Z.~Q.}\ \bibnamefont {Qiu}}, \bibinfo {author} {\bibfnamefont {R.~J.}\
  \bibnamefont {Cava}}, \bibinfo {author} {\bibfnamefont {S.~G.}\ \bibnamefont
  {Louie}}, \bibinfo {author} {\bibfnamefont {J.}~\bibnamefont {Xia}}, \ and\
  \bibinfo {author} {\bibfnamefont {X.}~\bibnamefont {Zhang}},\ }\href
  {\doibase 10.1038/nature22060} {\bibfield  {journal} {\bibinfo  {journal}
  {Nature}\ }\textbf {\bibinfo {volume} {546}},\ \bibinfo {pages} {265}
  (\bibinfo {year} {2017})}\BibitemShut {NoStop}%
\bibitem [{\citenamefont {Huang}\ \emph {et~al.}(2017)\citenamefont {Huang},
  \citenamefont {Clark}, \citenamefont {Navarro-Moratalla}, \citenamefont
  {Klein}, \citenamefont {Cheng}, \citenamefont {Seyler}, \citenamefont
  {Zhong}, \citenamefont {Schmidgall}, \citenamefont {McGuire}, \citenamefont
  {Cobden}, \citenamefont {Yao}, \citenamefont {Xiao}, \citenamefont
  {Jarillo-Herrero},\ and\ \citenamefont {Xu}}]{Huang2017}%
  \BibitemOpen
  \bibfield  {author} {\bibinfo {author} {\bibfnamefont {B.}~\bibnamefont
  {Huang}}, \bibinfo {author} {\bibfnamefont {G.}~\bibnamefont {Clark}},
  \bibinfo {author} {\bibfnamefont {E.}~\bibnamefont {Navarro-Moratalla}},
  \bibinfo {author} {\bibfnamefont {D.~R.}\ \bibnamefont {Klein}}, \bibinfo
  {author} {\bibfnamefont {R.}~\bibnamefont {Cheng}}, \bibinfo {author}
  {\bibfnamefont {K.~L.}\ \bibnamefont {Seyler}}, \bibinfo {author}
  {\bibfnamefont {D.}~\bibnamefont {Zhong}}, \bibinfo {author} {\bibfnamefont
  {E.}~\bibnamefont {Schmidgall}}, \bibinfo {author} {\bibfnamefont {M.~A.}\
  \bibnamefont {McGuire}}, \bibinfo {author} {\bibfnamefont {D.~H.}\
  \bibnamefont {Cobden}}, \bibinfo {author} {\bibfnamefont {W.}~\bibnamefont
  {Yao}}, \bibinfo {author} {\bibfnamefont {D.}~\bibnamefont {Xiao}}, \bibinfo
  {author} {\bibfnamefont {P.}~\bibnamefont {Jarillo-Herrero}}, \ and\ \bibinfo
  {author} {\bibfnamefont {X.}~\bibnamefont {Xu}},\ }\href {\doibase
  10.1038/nature22391} {\bibfield  {journal} {\bibinfo  {journal} {Nature}\
  }\textbf {\bibinfo {volume} {546}},\ \bibinfo {pages} {270} (\bibinfo {year}
  {2017})}\BibitemShut {NoStop}%
\bibitem [{\citenamefont {Miao}\ \emph {et~al.}(2018)\citenamefont {Miao},
  \citenamefont {Xu}, \citenamefont {Zhu}, \citenamefont {Zhou},\ and\
  \citenamefont {Sun}}]{Miao2018}%
  \BibitemOpen
  \bibfield  {author} {\bibinfo {author} {\bibfnamefont {N.}~\bibnamefont
  {Miao}}, \bibinfo {author} {\bibfnamefont {B.}~\bibnamefont {Xu}}, \bibinfo
  {author} {\bibfnamefont {L.}~\bibnamefont {Zhu}}, \bibinfo {author}
  {\bibfnamefont {J.}~\bibnamefont {Zhou}}, \ and\ \bibinfo {author}
  {\bibfnamefont {Z.}~\bibnamefont {Sun}},\ }\href {\doibase
  10.1021/jacs.7b12976} {\bibfield  {journal} {\bibinfo  {journal} {J. Am.
  Chem. Soc.}\ }\textbf {\bibinfo {volume} {140}},\ \bibinfo {pages} {2417}
  (\bibinfo {year} {2018})}\BibitemShut {NoStop}%
\bibitem [{\citenamefont {Chumak}\ \emph {et~al.}(2015)\citenamefont {Chumak},
  \citenamefont {Vasyuchka}, \citenamefont {Serga},\ and\ \citenamefont
  {Hillebrands}}]{Chumak2015}%
  \BibitemOpen
  \bibfield  {author} {\bibinfo {author} {\bibfnamefont {A.~V.}\ \bibnamefont
  {Chumak}}, \bibinfo {author} {\bibfnamefont {V.~I.}\ \bibnamefont
  {Vasyuchka}}, \bibinfo {author} {\bibfnamefont {A.~A.}\ \bibnamefont
  {Serga}}, \ and\ \bibinfo {author} {\bibfnamefont {B.}~\bibnamefont
  {Hillebrands}},\ }\href {\doibase 10.1038/nphys3347} {\bibfield  {journal}
  {\bibinfo  {journal} {Nat. Phys.}\ }\textbf {\bibinfo {volume} {11}},\
  \bibinfo {pages} {453} (\bibinfo {year} {2015})}\BibitemShut {NoStop}%
\bibitem [{\citenamefont {Chumak}\ and\ \citenamefont
  {Schultheiss}(2017)}]{Chumak2017}%
  \BibitemOpen
  \bibfield  {author} {\bibinfo {author} {\bibfnamefont {A.~V.}\ \bibnamefont
  {Chumak}}\ and\ \bibinfo {author} {\bibfnamefont {H.}~\bibnamefont
  {Schultheiss}},\ }\href {\doibase 10.1088/1361-6463/aa7715} {\bibfield
  {journal} {\bibinfo  {journal} {J. Phys. D. Appl. Phys.}\ }\textbf {\bibinfo
  {volume} {50}},\ \bibinfo {pages} {300201} (\bibinfo {year}
  {2017})}\BibitemShut {NoStop}%
\end{thebibliography}%

\end{document}